\documentclass{jpp}

\usepackage{graphicx}
\usepackage{mathrsfs}
\usepackage{amsmath}
\usepackage{natbib}
\usepackage{color}
\usepackage{multirow}
\usepackage[dvipsnames]{xcolor}

\usepackage{changes}
\definechangesauthor[color=blue]{LF}
\definechangesauthor[color=red]{SSC}

\ifCUPmtlplainloaded \else
  \checkfont{eurm10}
  \iffontfound
    \IfFileExists{upmath.sty}
      {\typeout{^^JFound AMS Euler Roman fonts on the system,
                   using the 'upmath' package.^^J}%
       \usepackage{upmath}}
      {\typeout{^^JFound AMS Euler Roman fonts on the system, but you
                   dont seem to have the}%
       \typeout{'upmath' package installed. JPP.cls can take advantage
                 of these fonts, if you use 'upmath' package.^^J}%
      }
  \else
  \fi
\fi

\ifCUPmtlplainloaded \else
  \checkfont{msam10}
  \iffontfound
    \IfFileExists{amssymb.sty}
      {\typeout{^^JFound AMS Symbol fonts on the system, using the
                'amssymb' package.^^J}%
       \usepackage{amssymb}%
         \let\leq=\leqslant
         \let\geq=\geqslant
      }{}
  \fi
\fi

\ifCUPmtlplainloaded \else
  \IfFileExists{amsbsy.sty}
    {\typeout{^^JFound the 'amsbsy' package on the system, using it.^^J}%
     \usepackage{amsbsy}}
    {\providecommand\boldsymbol[1]{\mbox{\boldmath $##1$}}}
\fi

\providecommand\bnabla{\boldsymbol{\nabla}}

\newsavebox{\astrutbox}
\sbox{\astrutbox}{\rule[-5pt]{0pt}{20pt}}

\newcommand\Fv{{\bf F}}
\newcommand\Ev{{\bf E}}
\newcommand\Bv{{\bf B}}
\newcommand\dBv{\delta{\bf B}}
\newcommand\duv{\delta{\bf u}}
\newcommand\Jv{{\bf J}}

\newcommand\uv{{\bf u}}

\newcommand\ev{{\bf e}}

\newcommand\kv{{\bf k}}
\newcommand\rv{{\bf r}}
\newcommand\xv{{\bf x}}
\newcommand\vv{{\bf v}}

\newcommand\Omegaci{\Omega_{c{\rm i}}}

\newcommand\mi{m_{\rm i}}
\newcommand\di{d_{\rm i}}

\newcommand\rhoi{\rho_{\rm i}}

\newcommand\Tipara{T_{\|{\rm i}}}
\newcommand\Tiperp{T_{\perp{\rm i}}}

\newcommand\vthi{v_{{\rm th,i}}}

\newcommand\betae{\beta_{\rm e}}
\newcommand\betai{\beta_{\rm i}}

\newcommand\rms{\rm rms}

\title[PIC and Eulerian hybrid-kinetic plasma turbulence]{Plasma turbulence at ion scales: a comparison between PIC and Eulerian hybrid-kinetic approaches}

\author[Cerri et al.]{S.~S.~Cerri$^1$\thanks{Email address for correspondence: silvio.cerri@df.unipi.it}, L.~Franci$^{2,3}$\thanks{Email address for correspondence: franci@arcetri.astro.it}, F.~Califano$^1$, S.~Landi$^{2,4}$, P.~Hellinger$^5$}

\affiliation{$^1$Physics Department ``E. Fermi'', University of Pisa, Largo B. Pontecorvo 3, I-56127 Pisa, Italy\\[\affilskip]
$^2$Dipartimento di Fisica e Astronomia, Universit\'a degli Studi
di Firenze, Largo E. Fermi 2, I-50125 Firenze, Italy\\[\affilskip]
$^3$INFN - Sezione di Firenze, Via G. Sansone 1, I-50019 Sesto
F.no (Firenze), Italy\\[\affilskip]
$^4$INAF - Osservatorio Astrofisico di Arcetri, Largo E. Fermi
5, I-50125 Firenze, Italy\\[\affilskip]
$^5$Astronomical Institute,  CAS, Bocni II/1401,  CZ-14100
Prague, Czech Republic}

\pubyear{2017}
\volume{?}
\pagerange{?}
\date{20 Dec 2016; Accepted 6 Mar 2017}
\begin{document}

\maketitle

% \deleted[id=LF]{(SW)} 
% \replaced[id=LF]{between}{of} --> "between" replaces "of"
\begin{abstract}

Kinetic-range turbulence in magnetized plasmas and, in particular, in
the context of solar-wind turbulence has been
extensively investigated over the past decades via numerical
simulations.  Among others, one of the widely
adopted reduced plasma model is the so-called hybrid-kinetic model,
where the ions are fully kinetic and the electrons are treated as a
neutralizing (inertial or massless) fluid. Within the same model,
different numerical methods and/or approaches to turbulence
development have been employed. In the present work, we present a
comparison between two-dimensional hybrid-kinetic simulations of plasma turbulence
obtained with two complementary approaches spanning
about two decades in wavenumber - from MHD inertial range to
scales well below the ion gyroradius - with a state-of-the-art accuracy.
One approach employs hybrid particle-in-cell (HPIC) simulations of freely-decaying Alfv\'enic
turbulence, whereas the other consists of Eulerian hybrid Vlasov-Maxwell (HVM) simulations
of turbulence continuously driven with partially-compressible large-scale fluctuations.  
Despite the completely different initialization and injection/drive at large scales, 
the same properties of turbulent fluctuations at $k_\perp\rhoi\gtrsim1$ are observed.
The system indeed self-consistently ``reprocesses'' the turbulent fluctuations while they are cascading towards smaller and smaller scales, in a way which actually depends on the plasma beta parameter.
Small-scale turbulence has been found to be mainly populated by kinetic Alfv\'en wave (KAW) fluctuations for $\beta\geq1$, 
whereas KAW fluctuations are only sub-dominant for low-$\beta$. 
\end{abstract}

%\begin{PACS}
%Authors should not enter PACS codes directly on the manuscript, as
%these must be chosen during the online submission process and will
%then be added during the typesetting process (see
%http://www.aip.org/pacs/ for the full list of PACS codes)
%\end{PACS}

%\newpage
%\tableofcontents
%\newpage 

\section{Introduction}\label{sec:Intro}

Studies on kinetic turbulence in collisionless magnetized plasmas are today considered
a major research area, especially in the field of solar wind (SW) turbulence~\citep{BrunoCarboneLRSP2013}.  
These studies have been powered by the availability of increasingly
detailed and accurate in-situ satellite measurements and
by the impressive increase of computational
resources for performing direct numerical investigations. 
In particular, spacecraft measurements show
that SW turbulent spectra exhibit power-law scaling spanning several
decades in frequency, with a spectral break
around the proton kinetic scales~\citep{BalePRL2005,SahraouiPRL2009, AlexandrovaPRL2009, RobertsAPJ2013, BrunoAPJL2014}.
At large, magnetohydrodynamics (MHD) scales, the turbulent SW magnetic
fluctuations follow very closely a Kolmogorov-like energy spectrum 
with a $-5/3$ slope.  In the kinetic range between the ion and
the electron kinetic scales, usually referred to as the ``dissipation''
(or ``dispersion'') range, a steepening of the magnetic spectrum 
is observed, with a spectral index typically close to $-2.8$. 
Conversely, measurements of the
electric fluctuations in the same range show a
shallower energy spectrum that overcomes its magnetic counterpart as
soon as the ion kinetic scales are crossed, with a spectral index
roughly between $-0.3$ and $-1$.  Several studies have tried to provide 
an explanation for the observed slopes in the kinetic range, 
either theoretically~\citep{StawickiJGR2001, GaltierPOP2003, HowesJGRA2008,
  GaryJGRA2009, SchekochihinAPJS2009, BoldyrevAPJL2012,
  BoldyrevAPJ2013, BoldyrevAPJ2015, PassotAPJL2015} or by means of numerical
simulations adopting different plasma models~\citep{HowesPRL2008, HowesPRL2011, ShaikhMNRAS2009, ParasharPOP2010,ParasharPOP2011, ValentiniPRL2010,
  ServidioPRL2012, ServidioJPP2015, PassotEPJD2014, Franci_al_2015b,
  Franci_al_2015a, ToldPRL2015,SulemAPJ2016, CerriAPJL2016}.  
Furthermore, the same plasma model based on the solution of the Vlasov equation 
can be implemented using different numerical techniques, such as a particle-in-cell (PIC) or 
an Eulerian method~\citep{MatthewsJCP1994, ValentiniJCP2007}. 
Last but not least, one can employ a continuous energy injection mechanisms
by adding an external source in the equations, or focus on a decaying-turbulence
scenario by imposing large-amplitude initial fluctuations. 
The former is the optimal choice for reaching a durable turbulent steady state, 
although a similar condition can be achieved even in the latter case, 
for a duration of the order of tens of the initial eddy turnover time, 
provided that a proper initialization is employed~\citep[see, e.g.,][]{Franci_al_2015b,Franci_al_2015a}. 
In this context, two recent studies of kinetic plasma turbulence have focused on the role of the plasma $\beta$ parameter 
(i.e., the ratio between the thermal pressure of the plasma and the equivalent magnetic pressure), 
within the framework of a hybrid Vlasov-Maxwell model. 
One adopted an Eulerian approach and presented an analysis on the properties of the small-scale fluctuations
 of externally driven turbulence~\citep{CerriAPJL2016}, the other one investigated the effects of
$\beta$ on the ion-scale spectral break in the magnetic field 
spectra of freely-decaying turbulence using a Lagrangian approach~\citep{Franci_al_2016b}.  
Because of the completely different numerical and energy injection method, a question that naturally arises 
is whether or not, and how, these results agree. 
Indeed, a fundamental point to be addressed is the sensitivity of the kinetic cascade to very different 
large-scale conditions as well as to the numerical treatment.  
In this work, we present a qualitative and quantitative comparison between two-dimensional high-resolution hybrid-kinetic
simulations of plasma turbulence performed with the HVM and the HPIC CAMELIA codes. 
The starting point is given by the simulations  
recently presented in \cite{CerriAPJL2016} and \cite{Franci_al_2016b}, respectively. 
Since these two numerical studies investigated slightly different values of the plasma beta, 
three additional simulations were performed with the HPIC code in order to explore exactly the same $\beta$ values as in \cite{CerriAPJL2016}, 
while keeping the original setting of \cite{Franci_al_2016b}.
One of the major points of interest in this
comparison is given by the fact that these simulations were not originally designed for a benchmark, thus adopting 
very different initial setup, injection mechanisms and complementary approaches to achieve the turbulent state.
Being aware of the intrinsic 3D nature of plasma turbulence, we stress that our ``2.5D"-3V approach 
is able to retain important features characterizing the turbulent dynamics that is expected to develop in the full 3D case, in the presence
of a background magnetic field~\citep{KarimabadiSSR2013,KarimabadiPOP2013,ServidioJPP2015,WanPOP2016,LiAPJL2016}.

The remainder of this paper is organized as follows. In
Sec.~\ref{sec:Model} we introduce the hybrid-kinetic model equations
and approximations, along with the specific numerical implementations
and simulation setup employed by the HPIC code
(Sec.~\ref{subsec:CAMELIA}) and by the HVM code
(Sec.~\ref{subsec:HVM}). In Sec.~\ref{sec:NumRes} we present a
comparison of the numerical results. In particular, we focus our attention on the 
shapes of magnetic structures (Sec.~\ref{subsec:Bperpstruct}), on the energy spectra of turbulent
fluctuations (Sec.~\ref{subsec:spectra}), and on the relation between
density and parallel magnetic spectra as expected for kinetic Alfv\'en
wave (KAW) fluctuations (Sec.~\ref{subsec:ratios}). Finally, we
provide a summary and discussion of the conclusions arising from the comparison in Sec.~\ref{sec:Conclusions}.

\section{The hybrid-kinetic model}
\label{sec:Model}

We adopt a hybrid approximation of the full Vlasov-Maxwell system of
equations for a quasi-neutral plasma, in which ions are fully kinetic
and electrons are modeled as a neutralizing massless fluid through a
generalized Ohm's law~\citep{WinskeSSRv1985, MatthewsJCP1994,
  ValentiniJCP2007}.  The actual treatment of the ion kinetics depends
on the numerical approach adopted and will be described below, in
Secs.~\ref{subsec:CAMELIA}--\ref{subsec:HVM}.  In the following,
equations are normalized to the ion mass, $\mi$, the ion cyclotron
frequency, $\Omega_{ci}$, the Alfv\'en velocity, $v_A$, and the ion
skin depth, $\di = v_A / \Omega_{ci}$.  The electromagnetic fields are
coupled to the ions via the non-relativistic low-frequency limit of
the Maxwell's equations, i.e., the Faraday's
and the Amp\'ere's laws,
%%%%%%%%%%%%%%%%%%%%%%%%%%%%%%%%%%%%%%%
\begin{subequations}\label{eq:Faraday-Ampere}
 \begin{equation}
  \frac{\partial\,\Bv}{\partial t} \,=\,  -\,\nabla\times\Ev\,,
 \end{equation}
 \begin{equation}
  \nabla\times\Bv\,=\,\Jv\,,
 \end{equation}
\end{subequations} 
%%%%%%%%%%%%%%%%%%%%%%%%%%%%%%%%%%%%%%%
where the displacement current term has been neglected in the latter.
The electrons' response is modeled via the generalized Ohm's law,
%%%%%%%%%%%%%%%%%%%%%%%%%%%%%%%%%%%%%%%
\begin{equation} \label{eq:ohm}
 \Ev\, =\, -\,\uv\times\Bv\, +\, \frac{\Jv\times\Bv}{n}\, -\, \frac{\nabla P_e}{n}\, +\, \eta\Jv 
\end{equation}
%%%%%%%%%%%%%%%%%%%%%%%%%%%%%%%%%%%%%%%
where $\eta$ is the resistivity, the electron inertia terms have been
neglected and we assumed quasi-neutrality, i.e., $n_i \simeq n_e=n$. 
The number density, $n$, and the ion bulk velocity, $\uv$, are computed as
velocity-space moments of the ions' distribution. An isothermal
equation of state is assumed for the scalar electron pressure,
$P_e=nT_{0e}$, with a given electron to ion temperature ratio
$\tau\equiv T_{0e}/T_{0i}$ at $t=0$.

\subsection{Hybrid Particle-In-Cell (HPIC) simulations of freely-decaying turbulence}
\label{subsec:CAMELIA}

In the HPIC method the ion distribution function is modeled in terms of (macro-)particles following
the trajectories given by the equation of motion,
%%%%%%%%%%%%%%%%%%%%%%%%%%%%%%%%%%%%%%%
\begin{subequations}\label{eq:trajectories}
 \begin{equation}
  \frac{{\rm d}\,\xv_i}{{\rm d}t}\,=\,\vv_i\,,
 \end{equation}
 \begin{equation}
  \frac{{\rm d}\,\vv_i}{{\rm d}t}\,=\,\Ev\,+\,\vv_i\times\Bv\,,
 \end{equation}
\end{subequations} 
%%%%%%%%%%%%%%%%%%%%%%%%%%%%%%%%%%%%%%%
where $\xv_i$ is the position and $\vv_i$ the velocity.

The HPIC simulations presented here have been performed with the code
CAMELIA (Current Advance Method Et cycLIc leApfrog), where the ions
are advanced by a Boris' scheme with an excellent long term accuracy.
The 2D computational domain lies in the $(x,y)$ plane and consists of
a $2048^2$ square box with resolution ${\rm d}x={\rm d}y = \di/8$ and
length $L=256\,\di$. We set a background magnetic field perpendicular
to the simulation plane, $\Bv_0=B_0\ev_z$, with $B_0=1$.  Accordingly,
each field $\Psi$ will be decomposed in its perpendicular (in-plane)
component, $\Psi_\perp$, and its parallel (out-of-plane, along
$\ev_{z}$) component, $\Psi_{\parallel}$, with respect to $\Bv_0$.  

We initialize with a spectrum of large-scale, in-plane, magnetic and
bulk velocity fluctuations, composed of a large 
number of Fourier modes with random phases and associated wavevectors 
in a range of almost a decade, between $k_{\perp,0}\,\di = 0.03$ 
and $0.28$. Such fluctuations are characterized by energy 
equipartition and vanishing correlation 
and their initial global amplitude is set to $\dBv_\perp^{\rms} =
\duv_\perp^{\rms} \sim 0.24$. The initial power 
spectrum of these fluctuations is proportional to $k_\perp$, so 
that more energy is contained in modes with larger wavevectors. 
Consequently, the higher modes have shorter associated eddy turnover 
time and are the first to contribute to the development of the
turbulent cascade. The lower modes act as an energy reservoir that
keeps feeding the cascade even after the maximum turbulent activity is
reached~\citep{Franci_al_2015b,Franci_al_2015a}, allowing the system to maintain a
quasy-steady state for a time of many eddy turnover times.  
Initially, we assume a uniform number
density $n_0 = 1$ and an ion temperature anisotropy $A_{\rm i} =
\Tiperp / \Tipara = 1$.
This setup is exactly the same setup as the one employed in \cite{Franci_al_2015b,Franci_al_2015a,Franci_al_2016b}. 
For the purpose of the present comparison, three new simulations have been performed, 
exploring the same values of the ion plasma beta recently investigated in \cite{CerriAPJL2016}, 
i.e., $\betai = 0.2$, $1$, and $5$.  Electrons are isotropic, with $\betae = \betai$ for each run.
A different number of particle-per-cell (ppc) has been employed for
the three simulations, since the ppc-noise in the density and in the ion bulk velocity fluctuations 
is larger for larger $\betai$, the number of particles being equal \citep{Franci_al_2016b}.  
In particular, $8000$, $16000$, and $64000$ ppc have been used for $\betai = 0.2$, $1$, and
$5$, respectively. This make the HPIC simulations presented here one of the
most accurate of this kind in the literature, with the total number of
particles in the whole grid reaching $\sim 2.7 \times 10^{11}$.

A non-zero resistivity has been introduced in order to guarantee a
satisfactory conservation of the total energy, with no claim to model
any realistic physical process. Its value has been chosen to be
$\eta=5 \times 10^{-4}$, in units of $4\pi \omega_p^{-1}$, based on
the discussion presented in \cite{Franci_al_2015b}, where different
values of $\eta$ were tested, and on the results of
\cite{Franci_al_2016b}, where simulations with the same setting (i.e.,
the same spatial resolution, injection scale, and amplitude of initial
fluctuations) and with many different values of $\beta$ were analyzed.

All the HPIC results shown here have been computed in correspondence
with the maximum turbulent activity. No energy injection by means of
external forcing is provided during the evolution, therefore turbulence, once developed, is freely decaying. 
However, such decay is observed to be quite slow and self-similar, so that the
spectral properties remain quite stable afterwards~\citep{Franci_al_2015b}.

\subsection{Hybrid Vlasov-Maxwell (HVM) simulations of externally-driven turbulence}\label{subsec:HVM}

The ions' dynamics in the HVM code consists of the forced Vlasov equation for the ion distribution function $f_i = f_i(\xv,\vv,t)$~\citep{CerriAPJL2016},
%%%%%%%%%%%%%%%%%%%%%%%%%%%%%%%%%%%%%%%
\begin{equation}\label{eq:vlasov_F}
 \frac{\partial\,f_i}{\partial t}\, + \vv\cdot\frac{\partial\,f_i}{\partial\xv}\, +\,
 \big(\Ev + \vv\times\Bv + \Fv_{ext}\big)\cdot\frac{\partial\,f_i}{\partial\vv} 
 \,=\,0\,,
\end{equation} 
%%%%%%%%%%%%%%%%%%%%%%%%%%%%%%%%%%%%%%%
where $\Fv_{ext}(\rv,t)$ is a $\delta$-correlated in time, external forcing which injects momentum in the system with a prescribed average power density $\varepsilon$. 
The external forcing has a correlation tensor that in Fourier space reads as
%%%%%%%%%%%%%%%%%%%%%%%%%%%%%%%%%%%%%%%
\begin{equation}\label{eq:Fcorrtens}
 \langle F_{\kv,i}(t)F_{\kv,j}^*(t')\rangle\,=\,
 \chi(k)\left[\alpha_1\left(1-\frac{k_ik_j}{|\kv|^2}\right)\,
 +\,\alpha_2\left(\frac{k_ik_j}{|\kv|^2}\right)\right]\delta(t-t')\,,
\end{equation}
%%%%%%%%%%%%%%%%%%%%%%%%%%%%%%%%%%%%%%%
where brackets denote ensemble averaging, $\kv$ is a wave vector and $\chi(k)$ is a scalar function depending on the modulus of the wavenumber only. The coefficients $\alpha_1$ and $\alpha_2$ quantify the relative degrees of incompressibility and compressibility of the forcing, respectively.
Eq.~\ref{eq:vlasov_F} is coupled with Eqs.~(\ref{eq:Faraday-Ampere})-(\ref{eq:ohm}), and such set of equations is solved on a fixed grid in multidimensional phase space, using an Eulerian algorithm which combines the so-called splitting scheme~\citep{ChengJCP1976,MangeneyJCP2002} with the current advance method (CAM)~\citep{MatthewsJCP1994}, explicitly adapted to the hybrid case~\citep{ValentiniJCP2007}. Here a 2D-3V phase space is considered (two dimensions in real space and three dimensions in velocity space). In order to avoid spurious numerical effects at very small scales, spectral filters which act only on the high-$k$ part of the spectrum are adopted~\citep{LeleJCP1992}. 

The two-dimensional real space is represented by a $L=20\pi\di$ squared box, with an uniform resolution ${\rm d}x={\rm d}y\simeq0.06\di$. The three-dimensional velocity domain is a cube limited by $-5\leq v/\vthi\leq+5$ in each direction with $51^3$ uniformly distributed points.
A check on velocity-space resolution has been carried out with $71^3$ grid points, showing no differences, 
and for the simulations presented here the conservation of the system's total mass and energy is satisfied 
with relative errors of the order of $10^{-3}$~\citep{ServidioAPJL2014}.
The initial condition is given by an uniform Maxwellian plasma, 
%%%%%%%%%%%%%%%%%%%%%%%%%%%%%%%%%%%%%%%
\[
f_{0i}(\vv)\,=\,\frac{n_0}{(2\pi\vthi^2)^{3/2}}\,e^{-|\vv|^2/(2\vthi^2)}
\]
%%%%%%%%%%%%%%%%%%%%%%%%%%%%%%%%%%%%%%%
with $n_0=1$ and $\vthi^2=\betai/2$, embedded in a constant background magnetic field perpendicular to the simulation plane, $\Bv_0=B_0\ev_z$ with $B_0=1$. Random small-amplitude 3D large-scale magnetic perturbations, $|\dBv(\rv)|\ll B_0$, with wave numbers in the range $0.1\leq (k_\perp\di)_{\dBv}\leq0.3$, are initially superposed to $\Bv_0$. Momentum injection is provided by the partially compressible external forcing, $\Fv_{ext}$, with $\alpha_1=\alpha_2=1/2$ and an average power input of $\varepsilon=5\times10^{-4}$. Such forcing acts at the smallest wave numbers of the system, $0.1\leq (k_\perp\di)_\Fv\leq0.2$, thus injecting energy only at the largest scales allowed by the simulation box. 
\\

\section{Numerical results}\label{sec:NumRes}

We present a comparison between three direct
numerical simulations performed with the HPIC code and three with
the HVM code. We stress that the two sets of simulations, in addition to the intrinsically
different numerical approach, implement very different initial conditions and a different way of
developing turbulence (cf. Secs.~\ref{subsec:CAMELIA}--\ref{subsec:HVM}). 
In particular, the HPIC simulations adopt Alfv\'enic-like large-amplitude initial magnetic and velocity perturbations,
$\dBv=\dBv_\perp$ and $\duv=\duv_\perp$, that freely decay into a fully turbulent state. 
On the other hand, the HVM simulations make use of a continuous external injection of 
partially-compressible momentum fluctuations, starting from generic 3D 
small-amplitude magnetic fluctuations and no initial velocity perturbations, 
$\dBv = \dBv_\perp + \delta B_\|\ev_z$ and $\duv=0$, until a quasi-steady turbulent state is reached.
Note that, even though the simulation is 2D, all vectors are three-dimensional, 
e.g. $\Bv(\rv)=B_x(x,y)\,\ev_x+B_y(x,y)\,\ev_y+B_z(x,y)\,\ev_z$. 
As explained above, both approaches allow the system to develop and maintain quite stable 
turbulent spectra for a time that is of the order of several eddy turnover times.
Consequently, the analysis of numerical results will be performed by considering a time average 
over a consistent part of such quasi-steady state for both the HPIC and the HVM simulations.
In all the simulations presented here, both HPIC and HVM, the temperature ratio is set to $\tau=1$. 
The same three initial values of the plasma beta are investigated, namely $\betai=0.2$, $1$ and $5$, 
letting us to explore the low-, intermediate-, and high-beta regimes, respectively. 
Although the initial fluctuations in HPC simulations fill a wider part of the MHD inertial spectrum (since the box size is larger) 
and have a much higher amplitude with respect to the HVM counterparts, the energy-containing scales are essentially 
the same for the two sets of simulations, i.e., $k_\perp\di\leq0.28$ and $k_\perp\di\leq0.3$, respectively. 
A summary of the different HPIC and HVM initialization is provided in Table~\ref{tab1} for a direct comparison.

%--------------------------------
\begin{table}
 \center
 \begin{tabular}{cccccccccc}
  \hline
    code & L[$\di$] & ${\rm d}x$[$\di$] & ppc \& $\vv$-space & $\delta B_{\perp,0}^{\rms}$ & $\delta B_{\|,0}^{\rms}$ & $(k_\perp\di)_{\dBv_0}$ & $\delta u_{\perp,0}^{\rms}$ & $\delta u_{\|,0}^{\rms}$ & $(k_\perp\di)_{\duv_0}$ \\
   \hline
    HPIC & 256 & 0.125 & 8000$\div$64000 & 0.24 & - & [0.03,0.28]& 0.24 & - & [0.03,0.28] \\
    & & & & & & & & &\\
    HVM & 20$\pi$ & 0.06 & $|\vv|\leq5\vthi$\, 51$^3$\,pts & 0.01 & 0.01 & [0.1,0.3] & - & - & -\\
  \hline
 \end{tabular}
   \caption{Synthetic comparison of differences between HPIC and HVM setup.}
   \label{tab1}
\end{table}
%--------------------------------

\subsection{Structures in real space}\label{subsec:Bperpstruct}

As a first step, we provide a qualitative picture of the
fully-developed turbulent dynamics arising in the two sets of simulations. 
A characteristic feature of turbulence in magnetized plasmas is
the formation of current sheets and coherent magnetic structures, 
as highlighted from either the early MHD and the more recent kinetic simulations~\citep[e.g.,][]{MatthaeusPOF1986,BiskampBOOK2003,ServidioNPG2011,ServidioPRL2012,KarimabadiPOP2013,Franci_al_2015b,NavarroPRL2016,CerriCalifanoNJP2017}, and recently observed also by direct measurements
in the solar wind~\citep{PerriPRL2012,ChasapisAPJL2015,GrecoAPJL2016}.
Also previous studies with HVM and with HPIC
have focused on kinetic effects related to magnetic structures~\citep{PerroneAPJ2013,ValentiniPOP2014,ValentiniNJP2016,FranciAIPCC2016}.
Therefore, an interesting quantity 
to look at is the modulus of the in-plane component
of the magnetic field, i.e. $|\Bv_\perp|\equiv\sqrt{B_x^2+B_y^2}$.
%--------------------------------
\begin{figure}
\advance\leftskip-0.2cm
\includegraphics[width=1.1\textwidth]{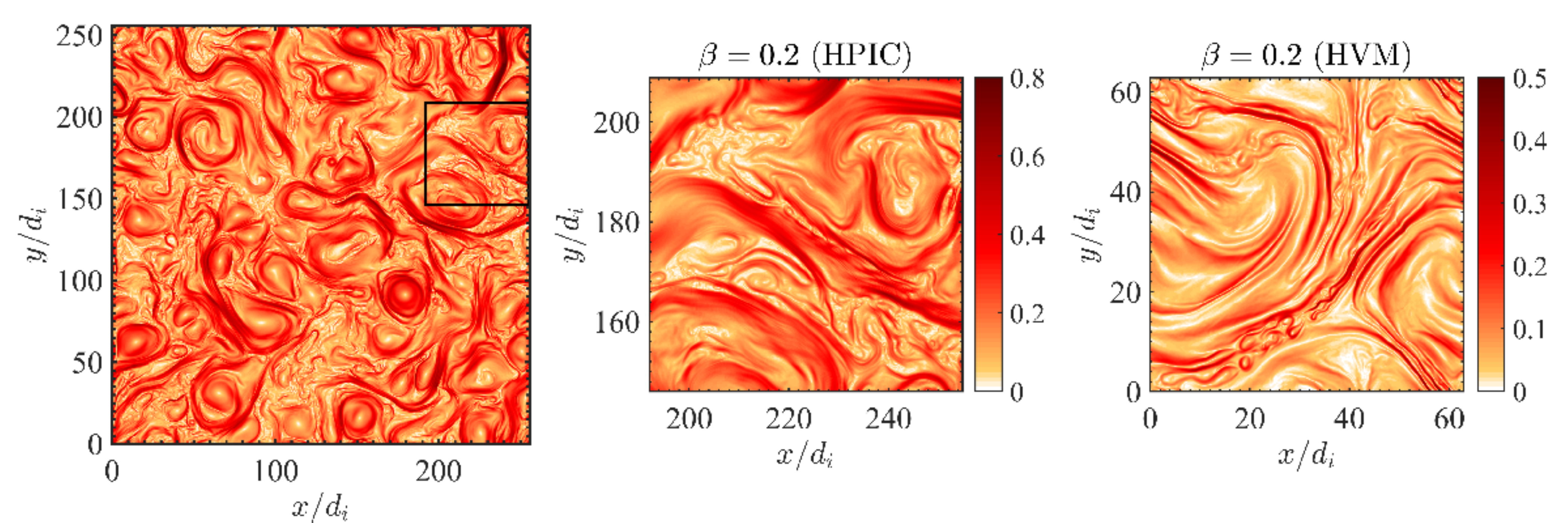}\\
\advance\leftskip-0.2cm
\includegraphics[width=1.1\textwidth]{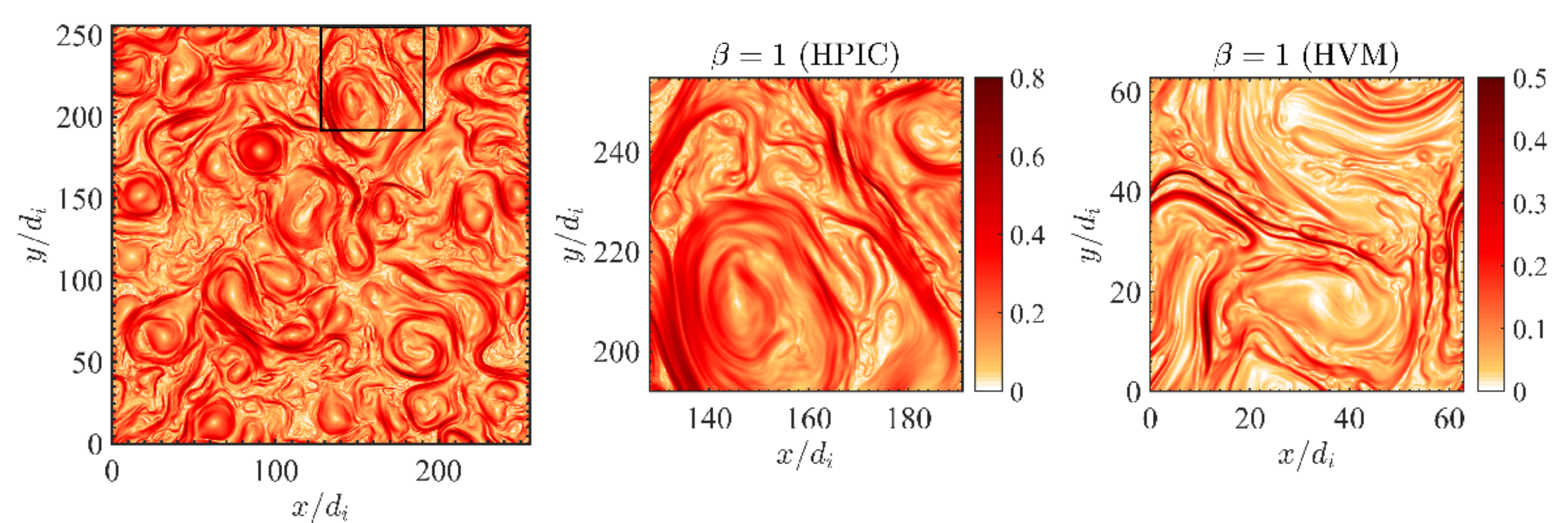}\\
\advance\leftskip-0.2cm
\includegraphics[width=1.1\textwidth]{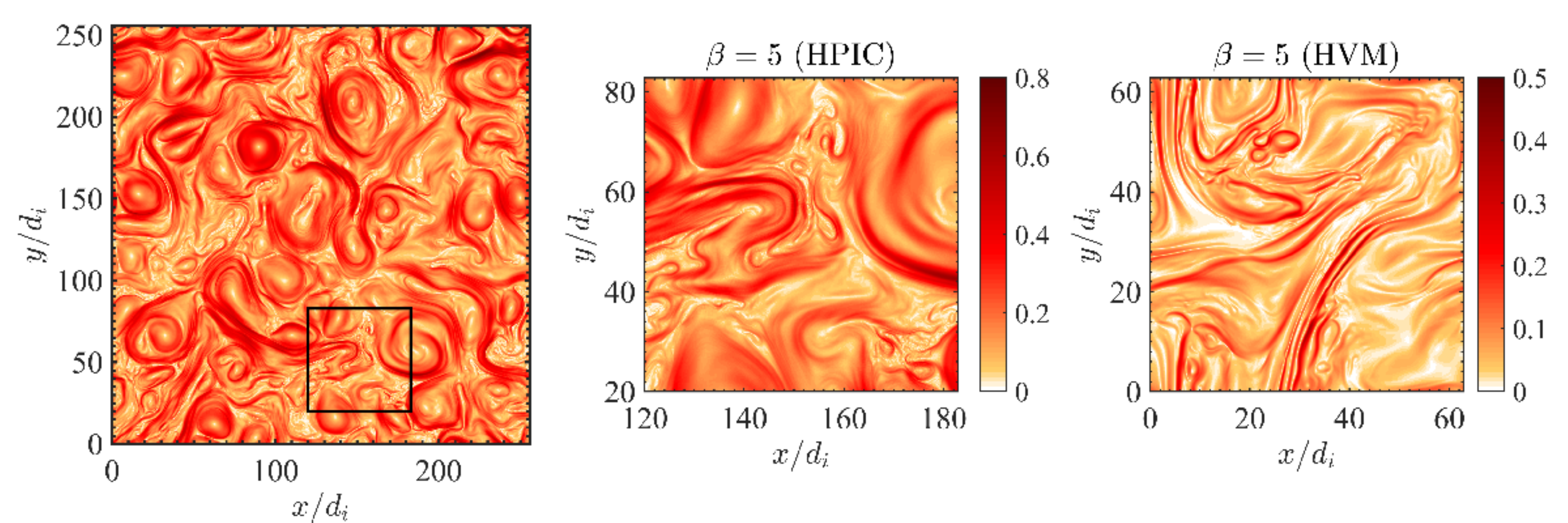}
\caption{Comparison between the in-plane magnetic field modulus, $|\mathrm{B}_{\perp}|$, in the HPIC and the HVM simulations, 
  for $\beta = 0.2$, $1$, and $5$ (\textit{top, middle and bottom row, respectively}).  
  We report either the entire HPIC boxes ({\it left column}) and a zoomed version of them ({\it middle column}), 
  in order to match the size of the HVM boxes ({\it right column}).}
\label{fig:structures}
\end{figure}
%--------------------------------
In Fig.~\ref{fig:structures}, we report the contour plots of $|\Bv_\perp|$ obtained with the HPIC and 
HVM simulations for the three values of the plasma beta, $\beta=0.2$ ({\it top row}), $1$ ({\it middle row}) 
and $5$ ({\it bottom row}). On the left frames, we draw the entire simulation boxes of the HPIC 
simulations, from which a zoom is shown in the central frames in order to match the size of the HVM simulation
boxes, which are instead shown on the right frames. The snapshots are taken at a give time, 
corresponding to the peak of the turbulent activity, $t_\textrm{peak}$, in the HPIC simulations and to a (random) 
time within the quasi-steady turbulent state in the HVM runs. 
Here $t_{\rm peak}$ is defined as the time at which the root-mean-square value 
of the current density, $\Jv$, presents a peak, thus indicating strong nonlinear (turbulent) activity~\citep{MininniPRE2009}.
Note that the difference in the initial rms-level of magnetic fluctuations between the two 
sets of simulations is quite large, i.e., a factor of $24$  when comparing only their perpendicular components and of about $15$ when also including the parallel components. While the level of fluctuations 
remain almost constant in the HPIC simulations (the relative decrease is less than $2\%$ between 
$t = 0$ and $t_\textrm{peak}$), they considerably increase, of about an order of magnitude, in HVM simulations. 
The differences in the total and perpendicular fluctuations reduce to a factor of $3$ and of $\sim 1.5$, 
respectively, during the quasi-steady phase. Consequently, both approaches achieve a value of the order 
of $\sim10\%$ of the initial background magnetic field (see Table~\ref{tab2}).
%--------------------------------
\begin{table}
 \center
 \begin{tabular}{cc|ccccccc|ccccccc}
  %\hline
     & & & \multicolumn{5}{c}{HPIC} & & & \multicolumn{5}{c}{HVM} & \\
     & & & & & & & & & & & & & & & \\
     & & & $\beta=0.2$ & & $\beta=1$ & & $\beta=5$ & & & $\beta=0.2$ & & $\beta=1$ & & $\beta=5$ & \\
   \hline
     $\delta B_\perp^{\rms}$ & & & 0.24 & & 0.23 & & 0.22 & & & 0.1 & & 0.1 & & 0.08 & \\
     & & & & & & & & & & & & & & & \\
     $\delta B_\|^{\rms}$    & & & 0.09 & & 0.08 & & 0.08 & & & 0.14 & & 0.14 & & 0.09 & \\
     & & & & & & & & & & & & & & & \\
     $\delta u_\perp^{\rms}$ & & & 0.18 & & 0.17 & & 0.14 & & & 0.33 & & 0.33 & & 0.35 & \\
     & & & & & & & & & & & & & & & \\
     $\delta u_\|^{\rms}$    & & & 0.07 & & 0.06 & & 0.07 & & & 0.02 & & 0.02 & & 0.02 & \\
     & & & & & & & & & & & & & & & \\
     $\delta n^{\rms}$       & & & 0.11 & & 0.08 & & 0.04 & & & 0.16 & & 0.14 & & 0.08 & \\
  \hline
 \end{tabular}
   \caption{Root-mean-square value of the fluctuations in the developed turbulent state.}
   \label{tab2}
\end{table}
%--------------------------------

A difference is observed in the early evolution of the simulations: with respect to the time at
which a fully-developed turbulent state is reached, the HPIC runs start developing current sheets and small-scale magnetic structures 
quite earlier than their HVM counterparts. This different  behavior is due to the different initialization, i.e., the very different level of initial
fluctuations and the very different number of modes, and to the different approach implemented for developing and sustaining
the turbulent cascade (free decay vs. forcing). In the HVM runs, the initial very small level of fluctuations increases due to the 
continuous energy injection (and the very long initial non-linear time decreases accordingly), thus determining a smooth transition 
from a weak- to a strong-turbulence regime. 
On the other hand, in the HPIC simulations, the injection-scale non-linear time at $t=0$ is already 
of the same order of its HVM counterpart in the quasi-steady turbulent state. 
As a result, in the HPIC case, many vortices and large-scale magnetic islands are suddenly generated and 
strong ion-scale gradients in the magnetic field (and, consequently, current sheets) quickly form between them.
Despite the different early evolution, it can be noted that once the turbulent cascade is fully developed, the two sets of simulations
exhibit the same qualitative behavior, for all the three beta cases, for what concerns the small-scale magnetic structures. 
In particular, all simulations exhibit reconnection occurring around the ion scales, 
leading to the formation of several small-scale island-like structures and 
to the full development of turbulence~\citep[see, e.g.,][]{CerriCalifanoNJP2017}.
These features can be indeed relevant in the context of the problem of 
turbulent dissipation in the solar wind as, for instance,
the so-called ``Turbulent Dissipation Challenge''~\citep{ParasharJPP2015}, 
where many observations have been focusing on the nature of magnetic fluctuations 
around the ion characteristic scales~\citep[see, e.g.,][and references therein]{RobertsJGRA2016,LionAPJ2016,PerroneAPJ2016}.
Coherent structures of larger sizes are clearly visible in the HPIC simulations, 
whereas they are much less evident in the HVM runs. 
In the HPIC simulations, the strong small-scale gradients are therefore embedded in a large-scale background 
with comparable energy. In the HVM simulations, instead, such large-scale background of $B_\perp$
fluctuations is much less energetic and almost all the visible magnetic structures exhibit a width of the the order of the
ion scales. These features of the two sets of simulations determine the observed sharper shapes and a higher contrast 
of the HVM contours in Fig.~\ref{fig:structures}.
  Such behavior is indeed
  even clearer when comparing the spectra of perpendicular and
  parallel magnetic fluctuations (see Fig.~\ref{fig:spectra_Bn}, right
  panels): an extremely good agreement is recovered across and below
  the ion kinetic scales, i.e. when any possible influence of the
  different initial setup and/or of the injection mechanism has faded
  away, whereas at the largest scales the two set of simulations
  dramatically differ, since the dominant contribution of magnetic
  fluctuations comes from the perpendicular component in HPIC
  simulations and from the parallel one in HVM simulations.  The
  behavior of magnetic structures in real space provide a first,
  qualitative, evidence that, as expected, the kinetic turbulent
  cascade and the consequent formation of small-scale structures
  essentially lose memory of the initial condition and/or of the
  injection-vs-decay mechanism, and they are relatively independent of
  the dissipation mechanisms. 

\subsection{Spectral properties of turbulent fluctuations}\label{subsec:spectra}

We now focus our attention on the energy spectra of the turbulent fluctuations. 
Note that for both the HPIC and the HVM simulations we show the spectra time-averaged over 
a time interval $\Delta t\simeq15 \, \Omegaci^{-1}$, corresponding to nearly half of the outer-scale nonlinear time. 
Since the rms-level of fluctuations at the injection scale and the numerical effects at the smallest scales
are different for the two set of simulations, we have rescaled the spectra of all quantities 
by a common factor in order to compare their behavior in the kinetic range at a given beta.
We have chosen such factor to be the ratio between the HVM and the HPIC total magnetic spectrum
at the latest scale before which the two intrinsically different numerical effects start to kick in, i.e. $k_\perp d_i=7$,
assumed to be the inner scale~\citep{BiskampBOOK2003}.
We stress that the applied shift is based only on the total magnetic spectrum, but in the kinetic range it will automatically 
produce overlapping spectra also for $\delta B_\|$, $\delta B_\perp$ and $\delta n$ (see below). 

In Fig.~\ref{fig:spectra_Bn}, we compare the power spectra of the density and total magnetic fluctuations, $E_n(k_\perp)$ and
$E_B(k_\perp)$ respectively ({\it left column}), and those of the parallel and perpendicular magnetic fluctuations separately,
$E_{B\|}(k_\perp)$ and $E_{B\perp}(k_\perp)$ respectively ({\it right column}). The comparison is shown for the three 
different values of the plasma beta, $\beta=0.2$ ({\it top row}), $1$ ({\it middle row}) and $5$ ({\it bottom row}). 
The wavenumber axis is given in $\di$ units, and a $\beta$-dependent $k_\perp\rhoi=1$ vertical line is displayed in the plots. 
The gray-shaded area {\bf at $k_\perp d_i>7$} highlights the part of the spectrum that is potentially affected by numerical effects, 
namely when the HPIC simulations are close to the ppc-noise level and when numerical filtering is significant in the HVM runs
(note that this is a conservative choice, since, for instance, density and magnetic field spectra typically exhibit a power law beyond $k_\perp d_i=7$).
%--------------------------------
\begin{figure}
\advance\leftskip-0.1cm
\includegraphics[width=0.51\textwidth]{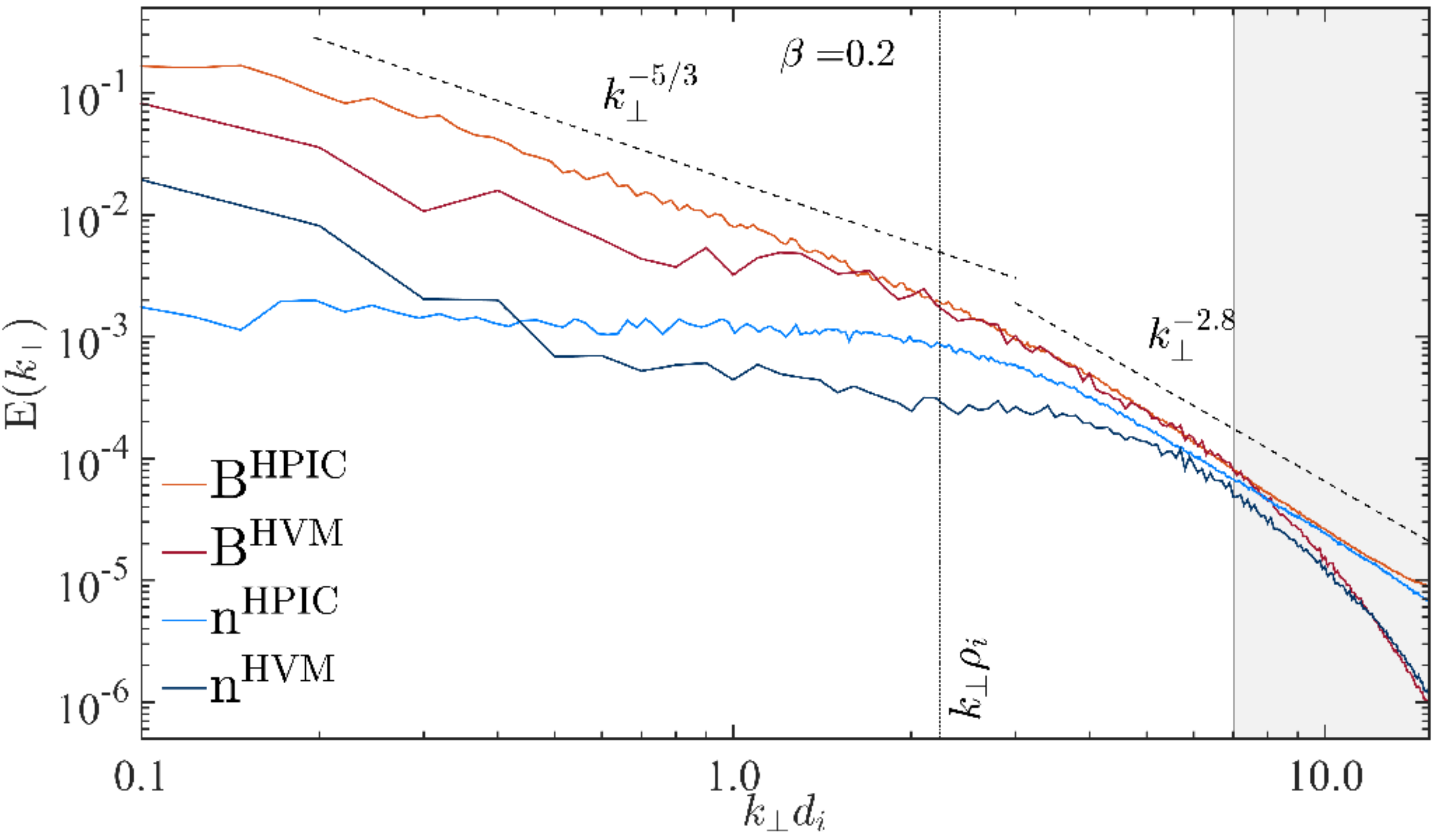}
\includegraphics[width=0.51\textwidth]{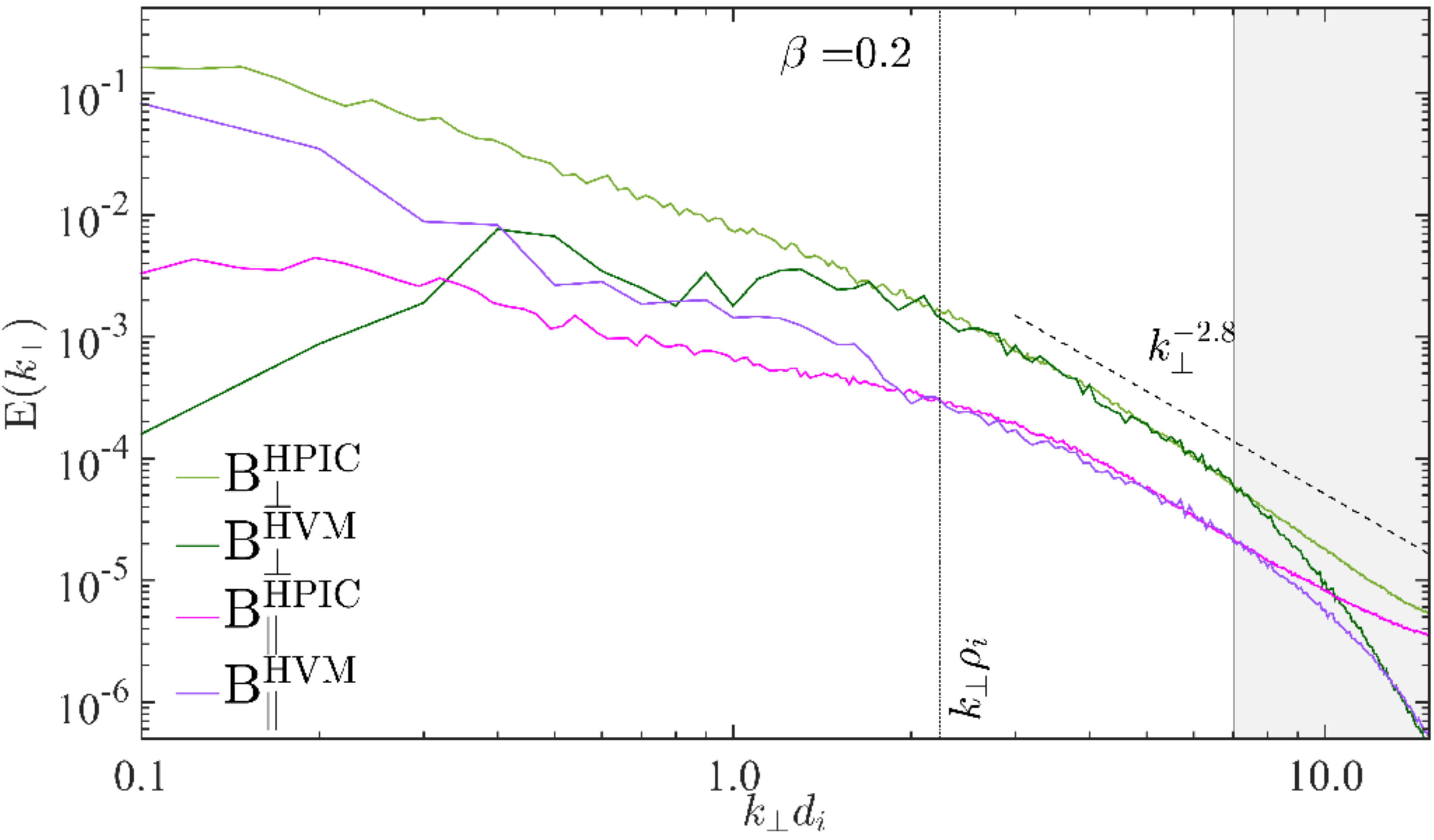}\\
\advance\leftskip-0.1cm
\includegraphics[width=0.51\textwidth]{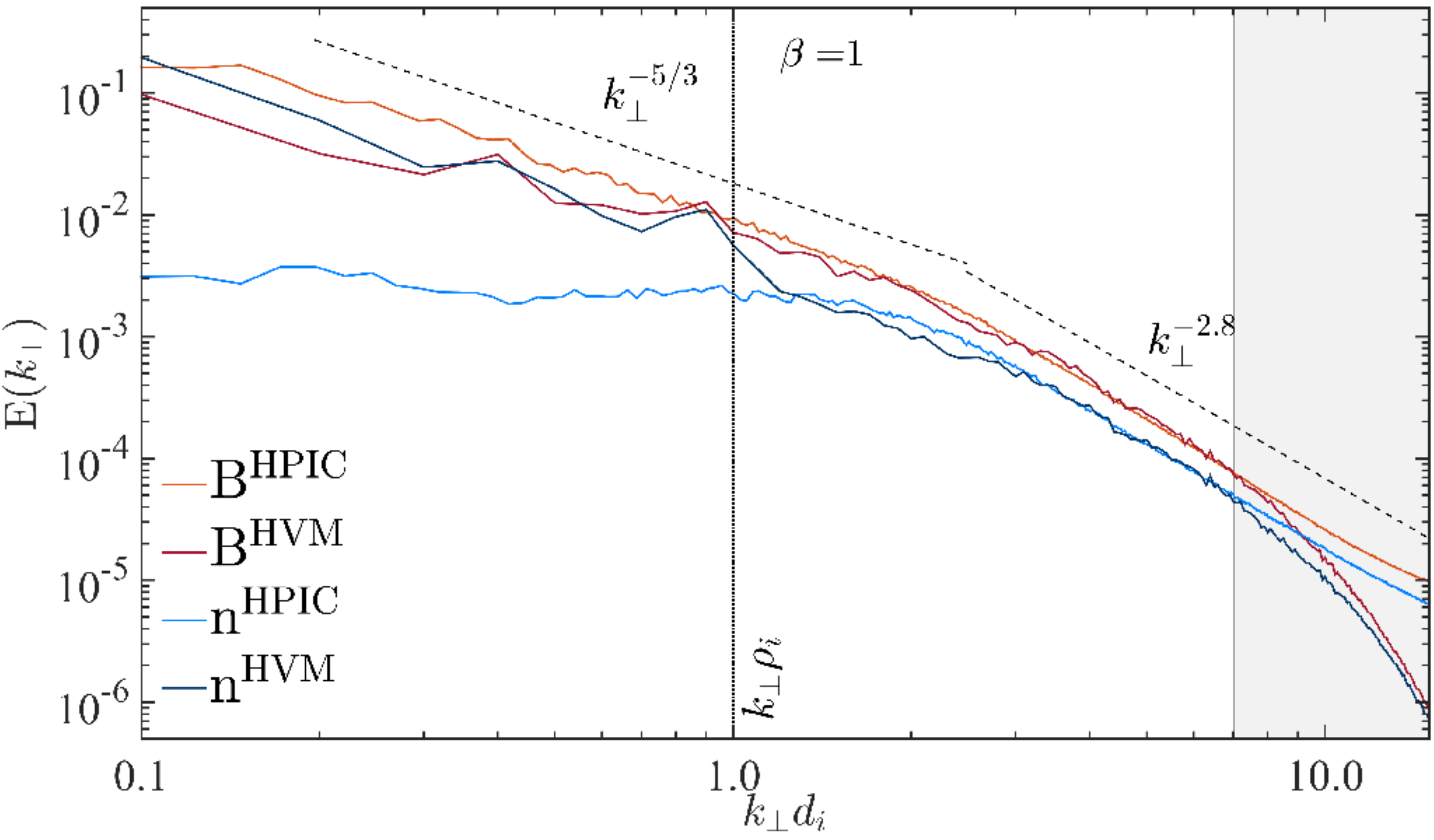}
\includegraphics[width=0.51\textwidth]{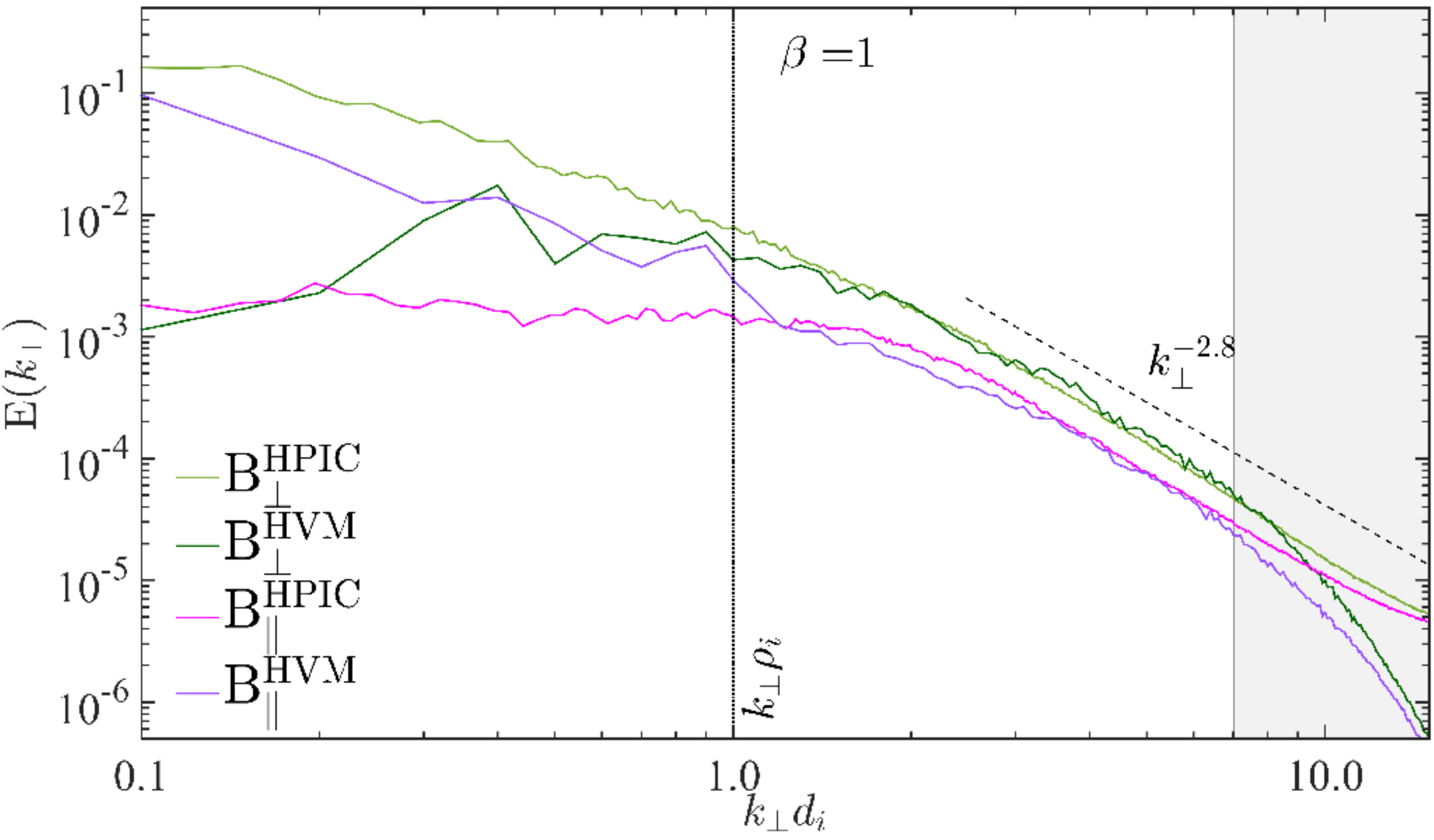}\\
\advance\leftskip-0.1cm
\includegraphics[width=0.51\textwidth]{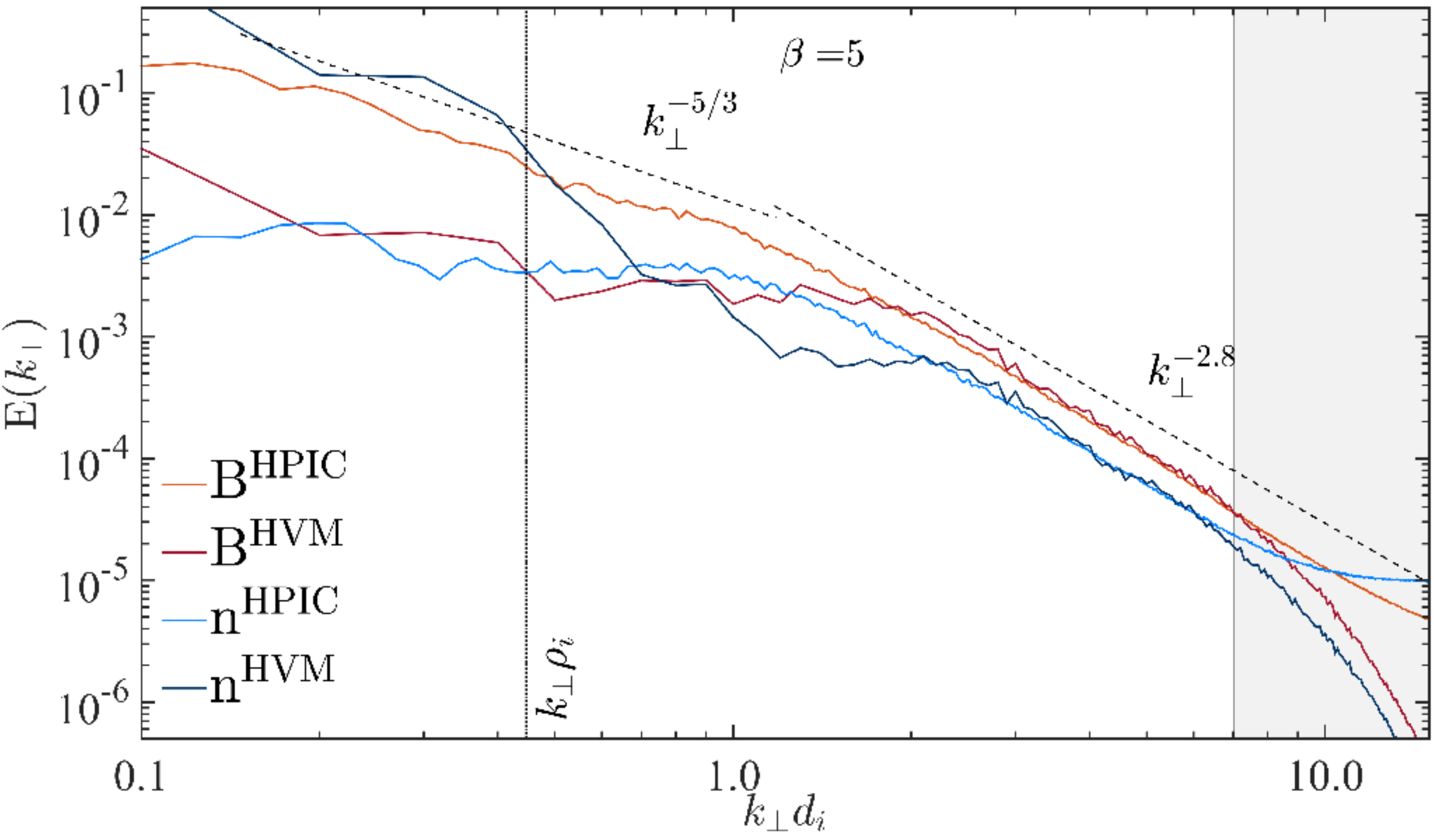}
\includegraphics[width=0.51\textwidth]{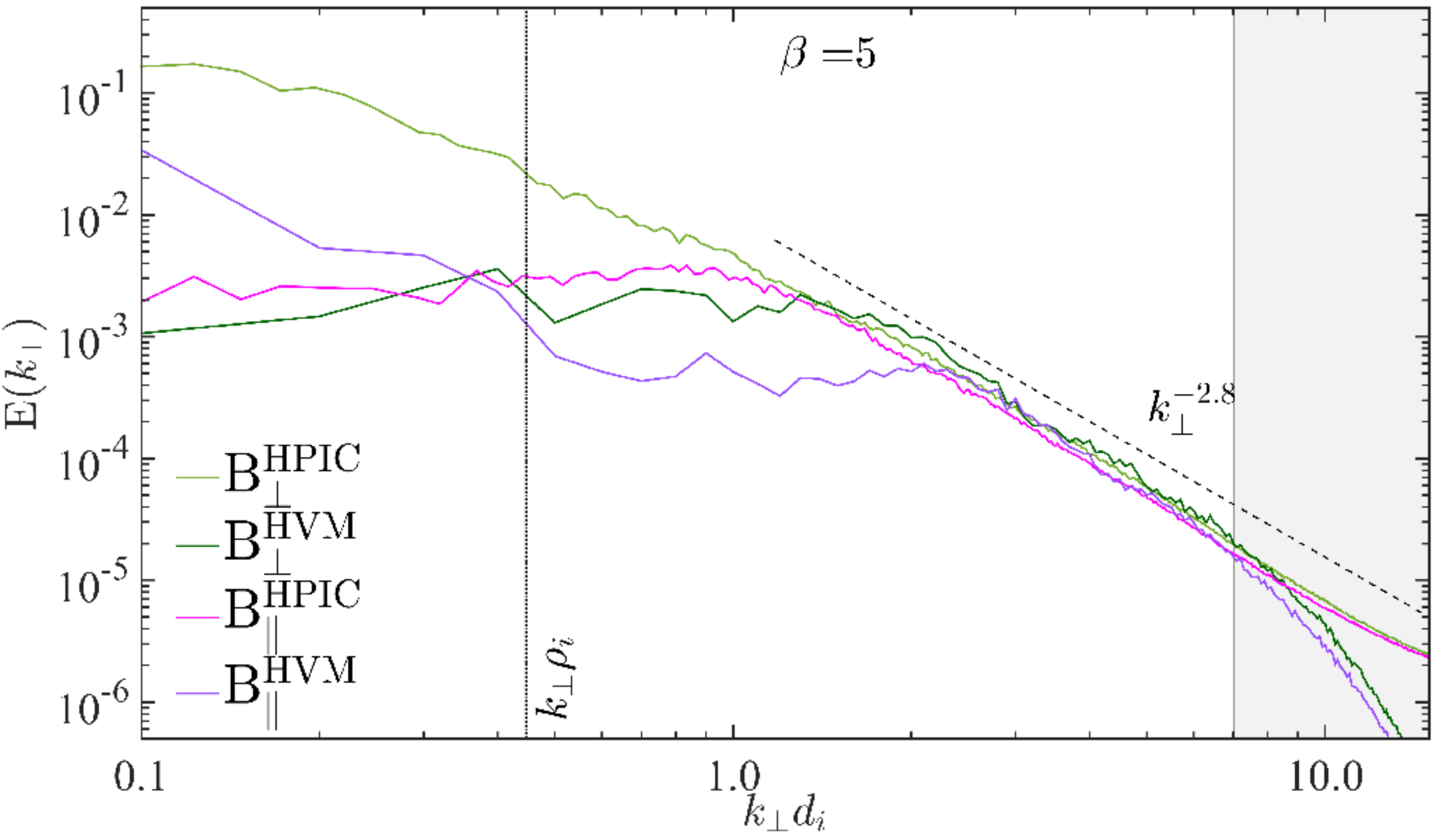}
\caption{Comparison between the results of the HPIC and the HVM
  approaches for the power spectra of the density, $\mathrm{n}$,
  of the magnetic field, $\mathrm{B}$ (\textit{first column}), and of
  its perpendicular, $\mathrm{B}_{\perp}$, and parallel,
  $\mathrm{B}_{\parallel}$, components (\textit{second column}) for
  $\beta = 0.2$, $1$, and $5$ (\textit{top, middle, and bottom row, respectively}). 
  The gray shaded region marks the range where the spectra are affected by numerical effects.}
\label{fig:spectra_Bn}
\end{figure}
%--------------------------------

We see that the spectra are quite in disagreement at large scales, due to the 
different initialization adopted by the HPIC and the HVM simulations. 
This is particularly evident when looking at the density spectrum, $E_n$, 
or when comparing the parallel and perpendicular magnetic fluctuation spectra, $E_{B\|}$ and $E_{B\perp}$. 
In fact, on the one hand, the HVM simulations drive partially compressive large-scale momentum fluctuations that 
develop a higher level of large-scale density and parallel magnetic fluctuations with respect to the HPIC counterparts. 
On the other hand, the HPIC simulations implement a higher level of large-scale perpendicular magnetic fluctuations, 
with respect to the level reached in the HVM turbulent state at the same large scales. 
Nevertheless the turbulent spectra in the two cases agree more and more as the cascade goes on
and energy is transferred towards smaller and smaller scales, eventually reaching a complete agreement at ion scales. In particular, as the ion kinetic scales are approached, 
we observe a switch in the level of the parallel and perpendicular
magnetic spectra for the HVM cases, thus denoting a self-consistent ``readjustment'' of the system 
at small scales (see right panels of Fig.~\ref{fig:spectra_Bn}). 
At $\beta=5$, being the level of the large-scale density fluctuations larger than the magnetic counterpart in the HVM case, 
one finds the same behavior, i.e., the density and magnetic fluctuations levels switch while the turbulent cascade proceeds towards 
small scales (bottom left panel of Fig.~\ref{fig:spectra_Bn}). 
It is worth noticing that the agreement between the HPIC and HVM results for the high-$\beta$ regime is only met at 
$k_\perp\rhoi\gg1$, since the external forcing employed in the Vlasov simulations acts very close to the ion gyroradius. 
Therefore, the system needs some ``cascade-time'' in order to self-consistently re-process the turbulent fluctuations, leading to an agreement between HVM and HPIC at somewhat smaller scales than $k_\perp\rhoi\sim1$.
Within the kinetic range, where the HVM and HPIC spectra nearly overlap, 
one can compute the slopes as the average between the two spectra.
With this method, the spectral slopes for the total, parallel and perpendicular magnetic fluctuations 
turn out to be $\alpha_B\simeq-2.85$, $\alpha_{B\|}\simeq-2.8$ and $\alpha_{B\perp}\simeq-2.9$, 
respectively, for $\beta=1$. 
Analogously, at $\beta=0.2$ and $\beta=5$, the corresponding spectral slopes are instead
$\alpha_B\simeq-3$, $\alpha_{B\|}\simeq-2.6$, $\alpha_{B\perp}\simeq-3.1$, and 
$\alpha_B\simeq-2.9$, $\alpha_{B\|}\simeq-2.75$, $\alpha_{B\perp}\simeq-2.9$, respectively.
The density fluctuations, for those cases where a power-law can be identified, 
always set up a kinetic spectrum with a slope consistent with $\alpha_n\approx-2.8$.
These trends are consistent with the results presented in \citet{Franci_al_2016b}, 
where an accurate fitting procedure was employed and for a wider beta range.

%--------------------------------
\begin{figure}
\advance\leftskip-0.1cm
\includegraphics[width=0.51\textwidth]{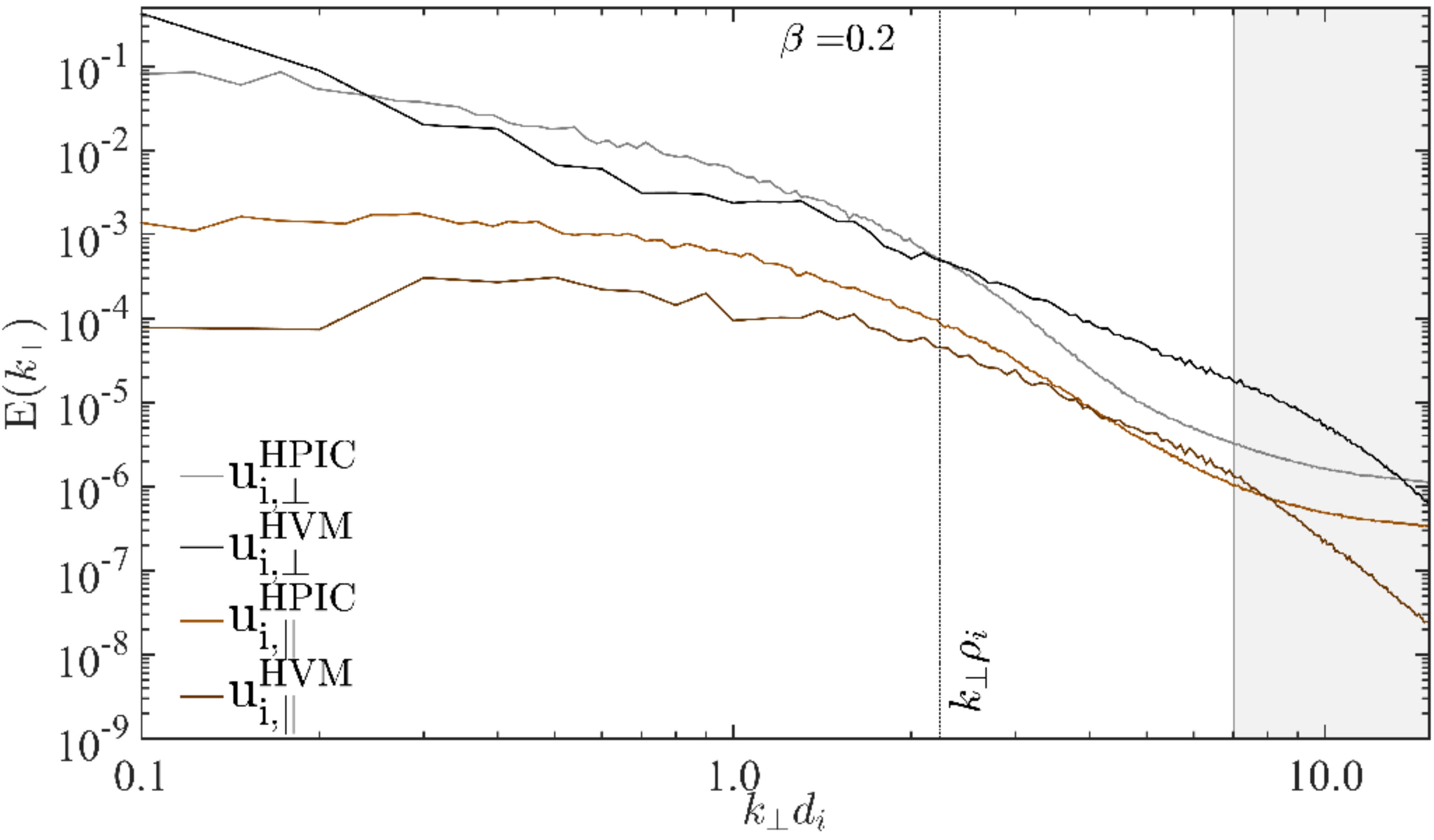}
\includegraphics[width=0.51\textwidth]{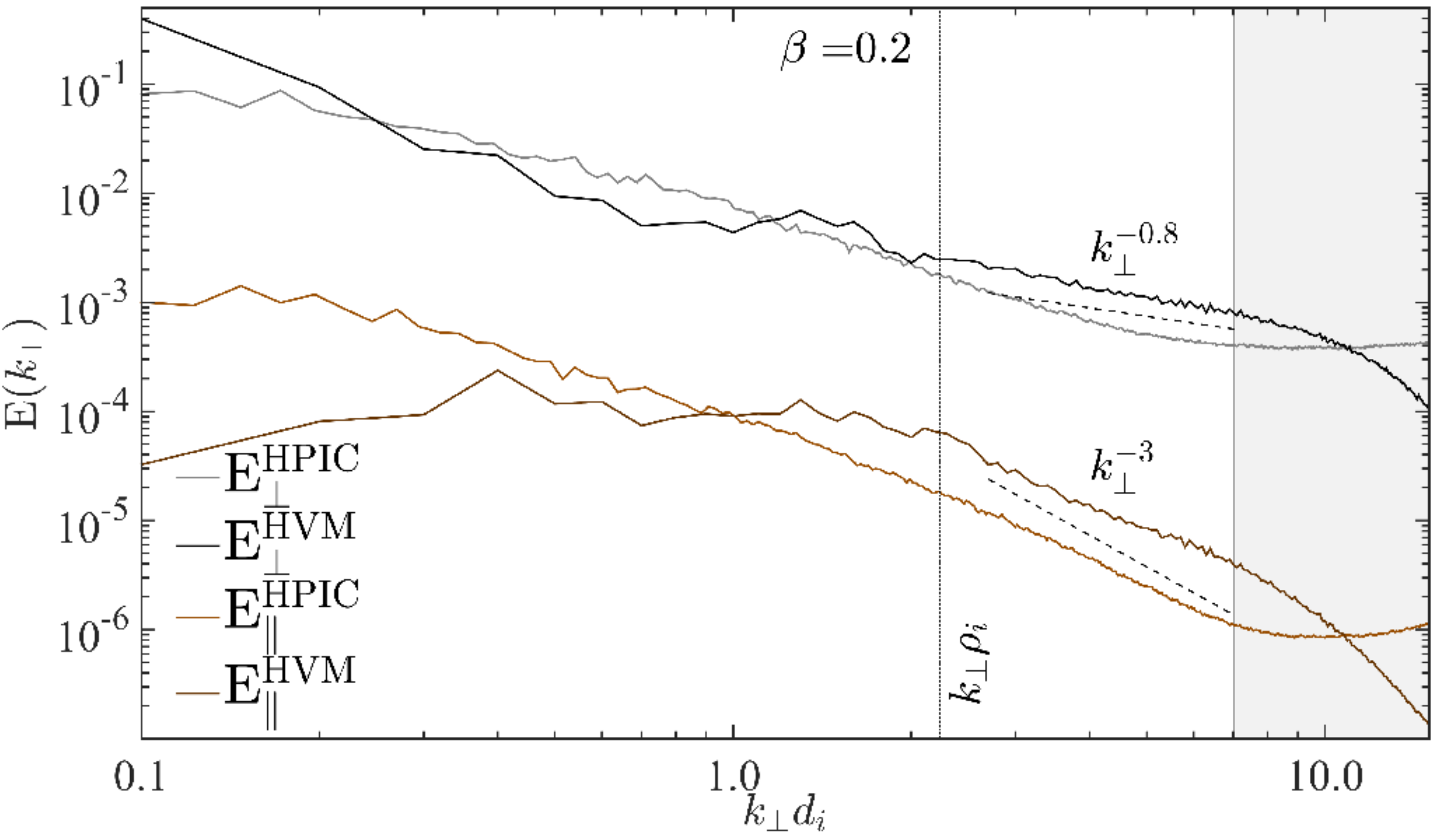}\\
\advance\leftskip-0.1cm
\includegraphics[width=0.51\textwidth]{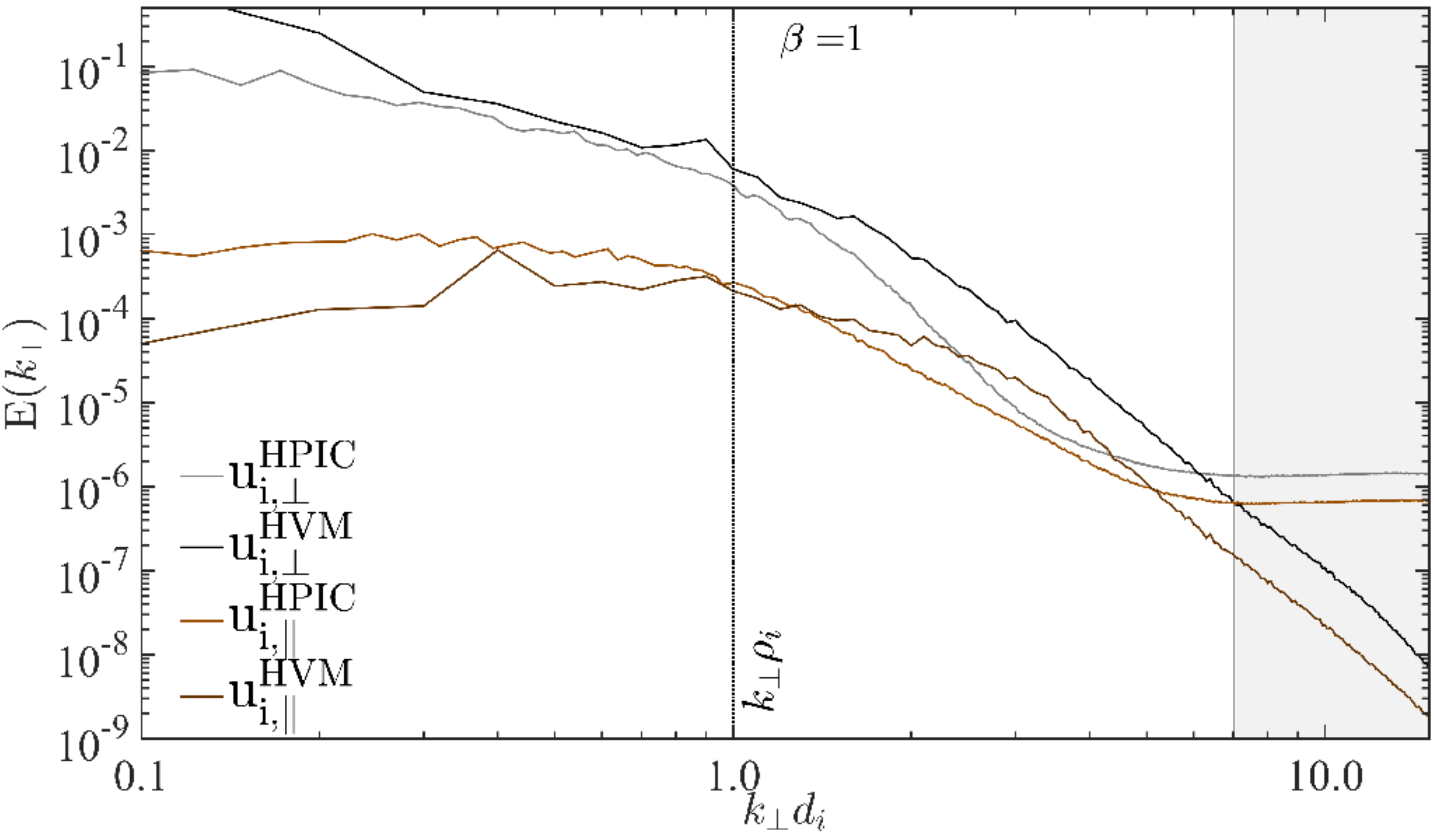}
\includegraphics[width=0.51\textwidth]{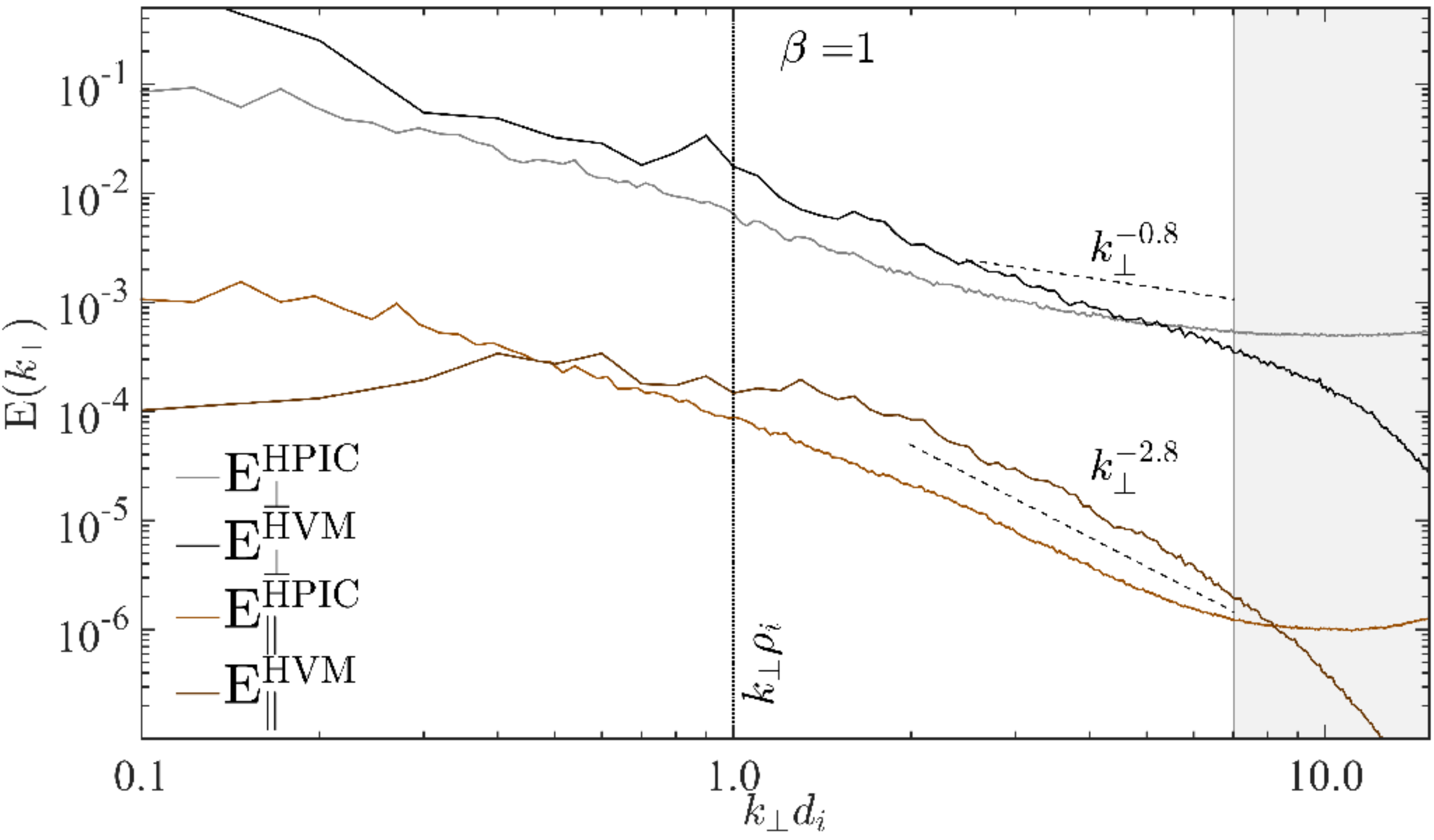}\\
\advance\leftskip-0.1cm
\includegraphics[width=0.51\textwidth]{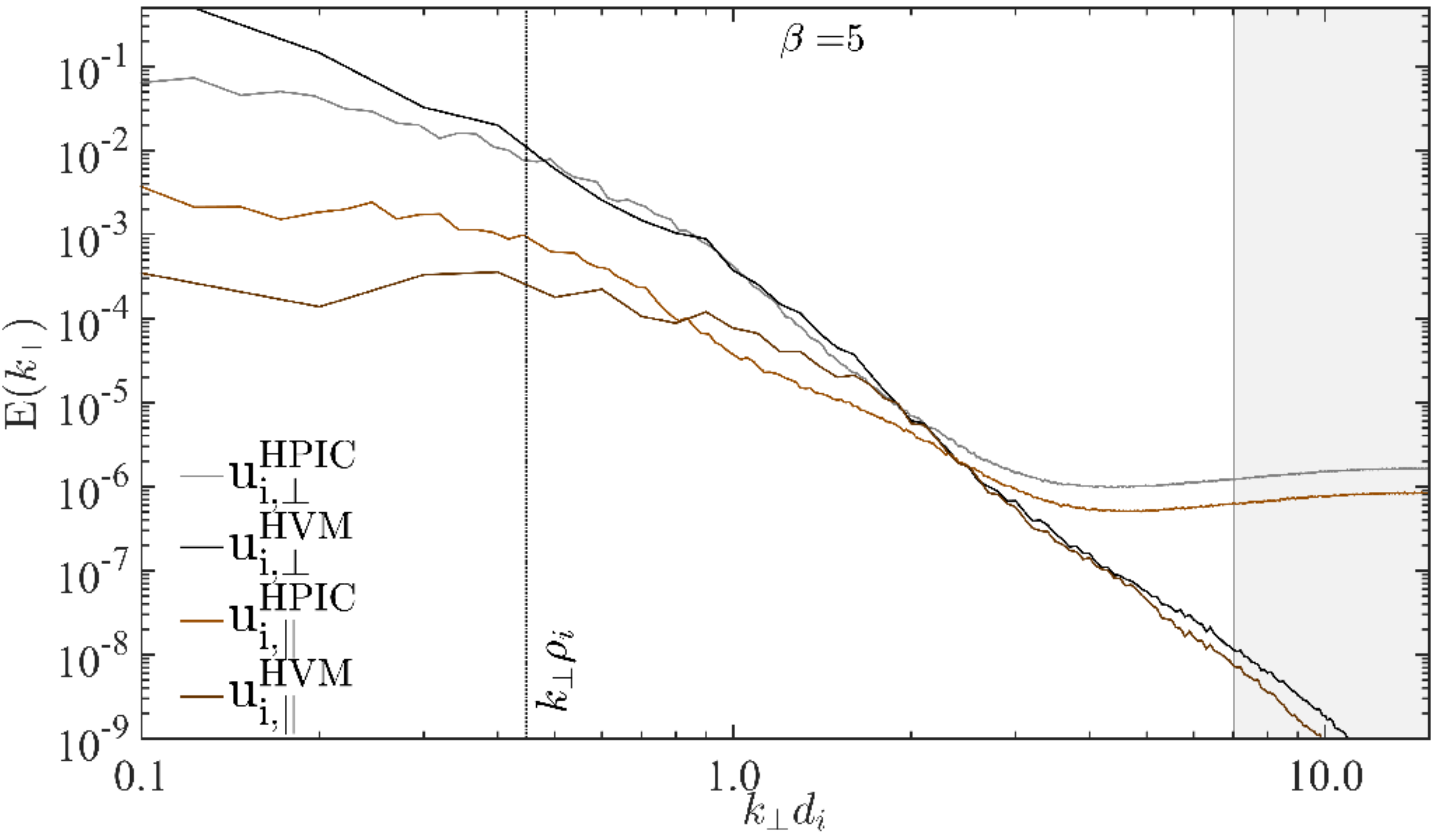}
\includegraphics[width=0.51\textwidth]{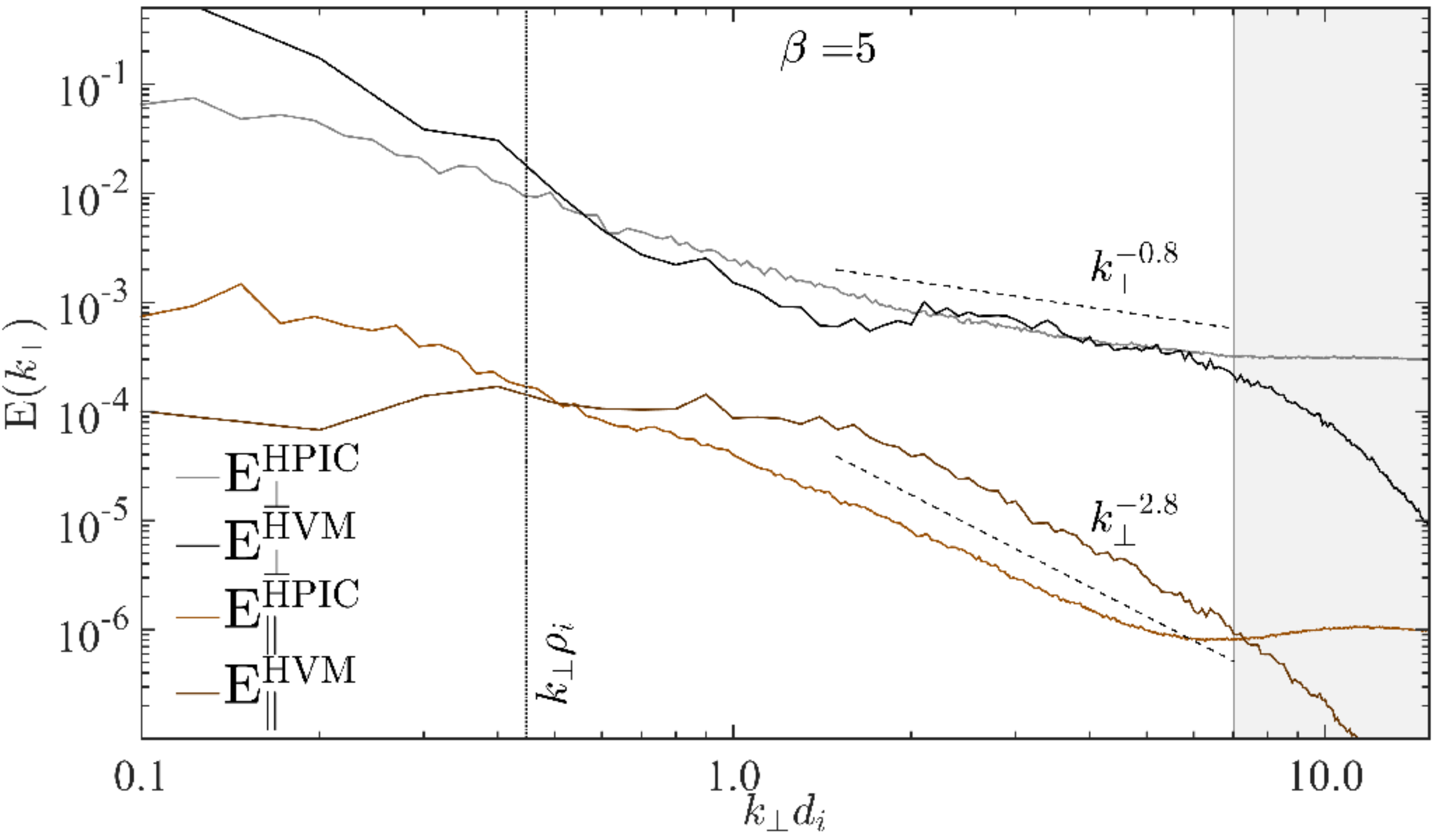}
\caption{The same as in Fig.~\ref{fig:spectra_Bn}, but for the power
  spectra of the perpendicular, $\mathrm{u}_{\mathrm{i},\perp}$, and
  parallel, $\mathrm{u}_{\mathrm{}i,\parallel}$, components of the ion
  bulk velocity (\textit{first column}) and of the perpendicular,
  $\mathrm{E}_{\perp}$, and parallel, $\mathrm{E}_{\parallel}$,
  components of the electric field (\textit{second column}).}
\label{fig:spectra_uE}
\end{figure}
%--------------------------------

Let us now focus our attention on the velocity and electric fluctuation spectra. 
These quantities are particularly sensitive to this comparison for two different reasons: 
first, the numerical treatment of the velocity space represents the main difference between HVM and HPIC, 
and, second, the electric field is a derived quantity and thus possibly more affected by 
the different numerical effects which are present in both codes.
In Fig.~\ref{fig:spectra_uE}, similarly to Fig.~\ref{fig:spectra_Bn}, we now compare the power spectra of the parallel and perpendicular
velocity fluctuations, $E_{u\|}(k_\perp)$ and $E_{u\perp}(k_\perp)$, respectively ({\it left column}), and the power spectra of the parallel and
perpendicular electric fluctuations, $E_{E\|}(k_\perp)$ and $E_{E\perp}(k_\perp)$, respectively ({\it right column}). 
Velocity fluctuations do indeed exhibit a quite significant disagreement between HPIC and HVM. 
In particular, at $\beta=0.2$ and $\beta=1$, it seems that the HVM spectra maintain a clear power law at smaller scales 
(before numerical damping starts to be effective, as indicated by the grey-shaded area), 
while their HPIC counterparts are damped as the ion-gyroradius scale is crossed, 
until a plateau is observed as the ppc-noise level is reached.
For $\beta=5$, instead, velocity spectra are steepening roughly in the same way at $k_\perp\rhoi>1$,
and thus a better agreement between the two methods is found.  
The different small-scale behavior of the velocity spectra for $\beta\leq 1$ does not seem to be a consequence of
the different approach (HPIC vs. HVM), but rather of the different injection method (free-decay vs. forcing). 
In this respect, it's worth mentioning that what is observed in the present HPIC simulations 
is rather comparable to what has been shown by previous HVM simulations of freely-decaying turbulence in \cite{ServidioJPP2015}. 
In fact, by looking at three cases with similar $\beta$ values in Fig.~4 of \cite{ServidioJPP2015}, 
we see that the velocity spectra exhibit a very similar damping, occurring at more or less the same scales. 
Therefore, the reason of such difference between the HPIC velocity spectra and the HVM counterparts presented here
is more likely due to a continuous injection of momentum in the HVM cases, which may be in general responsible for sustaining the cascade of velocity fluctuations (cf. also Table~\ref{tab2}).  
Finally, due to the quick drop of the HPIC velocity spectra at small scales, 
the ppc-noise level is reached slightly before $k_\perp d_i\sim 7$.
However note that the $y$-range used for these spectra is larger with respect to the one used for the other spectra
and thus, since their power is several orders of magnitude below the other fields, 
we do not expect that these features play a fundamental role in the other spectra (see Fig.~\ref{fig:spectra_Bn})
and in the Ohm's law at kinetic scale, where the MHD term related to $\uv$ is by definition negligible.
Also for what concerns the perpendicular and the parallel components of the electric field fluctuations, 
although $\Ev$ is a derived field (and thus more sensitive to numerics and noise due, e.g., to the density gradient),
we do observe a reasonable level of agreement between the HVM and the HPIC power spectra.
In this regard, since $\Ev$ is computed through the generalized Ohm's law, 
it is possible to recover a prediction for its slope in the kinetic range, by considering the contributions from its terms separately,
as previously done in \cite{Franci_al_2015a}.
The main contributions at sub-ion scales come from the Hall term, $\Ev^{\textrm{Hall}}$, and the electron pressure gradient term, 
$\Ev^{p\textrm{e}}$, since the steepening of the velocity spectra makes the MHD term, $\Ev^{\textrm{MHD}}$, negligible.
The leading terms of the perpendicular and of the parallel electric field at sub-ion scales are given by
%%%%%%%%%%%%%%%%%%%%%%%%%%%%%%%%%%%%%%%
\begin{subequations}
\begin{equation}
\Ev_\perp \sim \Ev_\perp^{\textrm{Hall}} + \Ev_\perp^{p\textrm{e}}
\propto (\kv_\perp \times \Bv_|) \times \Bv_0 - \bnabla \mathrm{p}_e \propto \bnabla 
(\Bv_0 \cdot \Bv_\parallel + T_e n )\,,
\end{equation}
\begin{equation}
\Ev_\| \sim \Ev_\|^{\textrm{Hall}} \propto (\kv_\perp \times \Bv_\|) \times \Bv_\perp,
\end{equation}
\end{subequations}
respectively. Consequently, the expected slope can be recovered as
\begin{subequations}
\begin{equation}
E_{E\perp} \propto k^2_\perp E_{B_\|,n},
\end{equation}
\begin{equation}
E_{E\|} \propto k^3_\perp E_{B_\|} E_{B_\perp}.
\end{equation}
\end{subequations}
%%%%%%%%%%%%%%%%%%%%%%%%%%%%%%%%%%%%%%%
Since the slope of $E_{B_\|}$ and $E_{B_n}$ at sub-ion scales is $\sim-2.8$ for all three beta regimes 
(see Fig.~\ref{fig:spectra_Bn}), the predicted slope of $E_{E_\perp}$ is $-0.8$ in all cases. 
On the contrary, the slope of $E_{B_\perp}$ slightly changes with $\beta$, 
being close to 
$-3.2$ for $\beta = 0.2$, $-3$ for $\beta = 1$, and $-3$ for $\beta = 5$ 
%\replaced[id=SSC]{$-2.8$ for $\beta = 0.2$, $-2.9$ for $\beta = 1$, and $-3$ for $\beta = 5$ }{
(Fig.~\ref{fig:spectra_Bn}, right panels). 
Therefore, the slope of $E_{E\|}$ is expected to be close to 
%$-2.8$ for $\beta =0.2$, $-2.7$ for $\beta = 1$, and $-2.6$ for $\beta = 5$.
%\replaced[id=SSC]{$-2.6$ for $\beta =0.2$, $-2.7$ for $\beta = 1$, and $-2.8$ for $\beta = 5$ }{
$-3.0$ for $\beta =0.2$, $-2.8$ for $\beta = 1$, and $-2.8$ for $\beta = 5$.
A very good agreement is observed between the results of all simulations and these
theoretical predictions for both the perpendicular and the parallel
electric spectra and for all three values of $\beta$, up to
$k_\perp\di\sim7$, where the HPIC spectra flattens due to numerical
noise while the HVM spectra drops due to filtering.
The only difference is observed in the spectrum of the perpendicular
electric field for $\beta = 1$, since the velocity spectra
of the HPIC and of the HVM runs start differing at $k_\perp\di \gtrsim 1$.
The larger level of fluctuations in the HVM case thus provides 
a non-negligible contribution of the MHD term, 
$\Ev_\perp^{\textrm{MHD}} \propto \uv_\perp \times \Bv_0$, at those scales,
which makes $E_{E_\perp}$ be a little steeper around ion scales.

\subsection{Relation between density and parallel magnetic fluctuations}\label{subsec:ratios}

By looking at the expected relation between the density and parallel magnetic fluctuations for KAW fluctuations, it was recently shown in
\cite{CerriAPJL2016} that turbulence properties in driven 2D3V HVM simulations were undergoing to a transition when passing from a
$\beta\geq1$ to a low-$\beta$ regime. 
In particular, at $\beta\geq1$, it was found that $\delta n$ and $\delta B_\|$ fluctuations were
satisfying the relation expected for KAW turbulence, whereas that was not the case for the low-$\beta$ regime.  
Here, we consider again this aspect of the small-scale turbulent fluctuations and we compare 
the HVM results with the HPIC simulations of freely-decaying large-scale Alfv\'enic fluctuations. 
The aim of this further comparison is to understand whether this transition depends on the particular
choice of the injection mechanism and on the nature of the large-scale fluctuations feeding the cascade,
or if there is a self-consistent re-processing of the large-scale turbulent fluctuations 
realized by the system as soon as the cascade crosses the ion kinetic scales.
%--------------------------------
\begin{figure}
\centering
\includegraphics[width=0.65\textwidth]{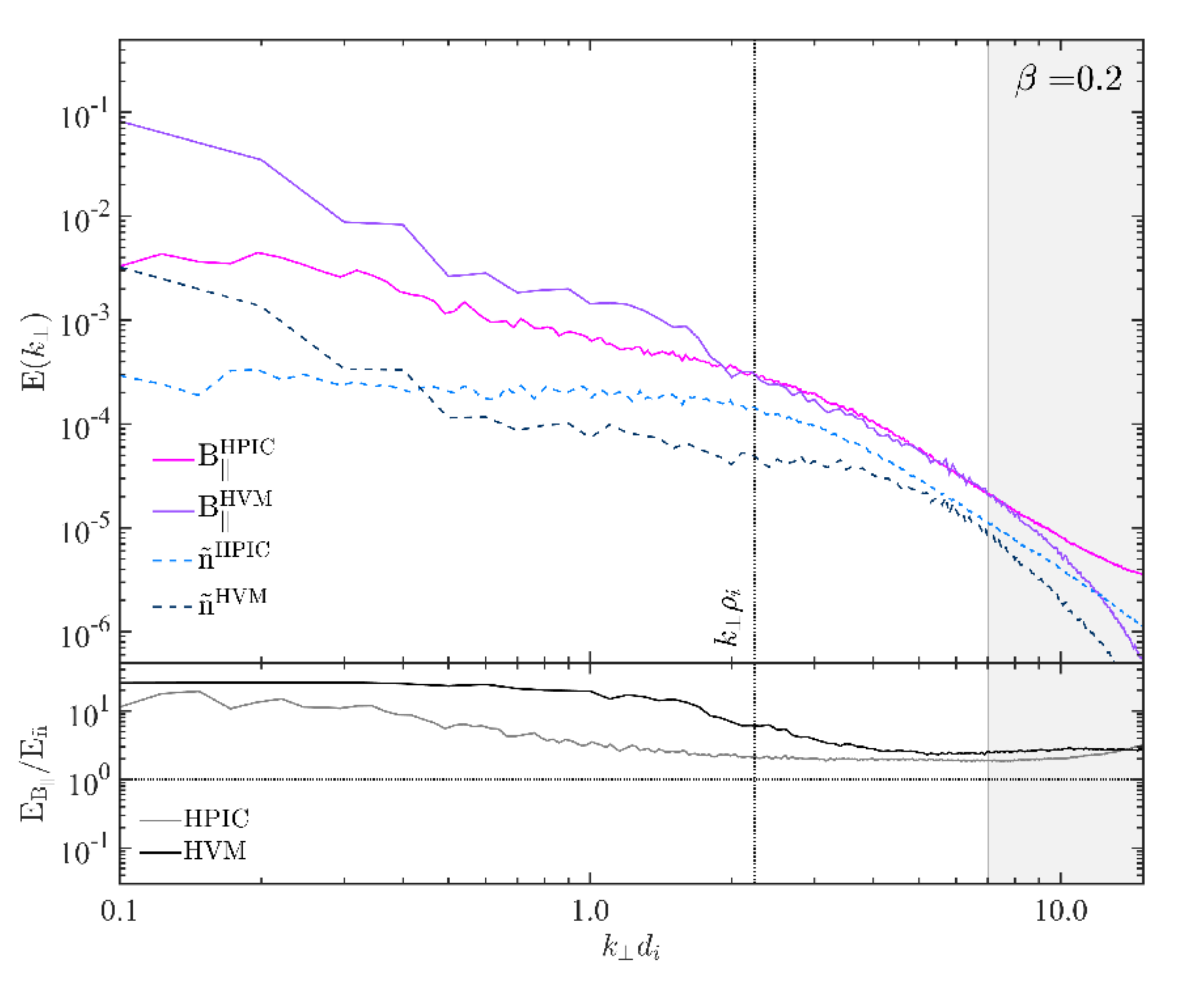}\\
\vspace{-0.35cm}
\includegraphics[width=0.65\textwidth]{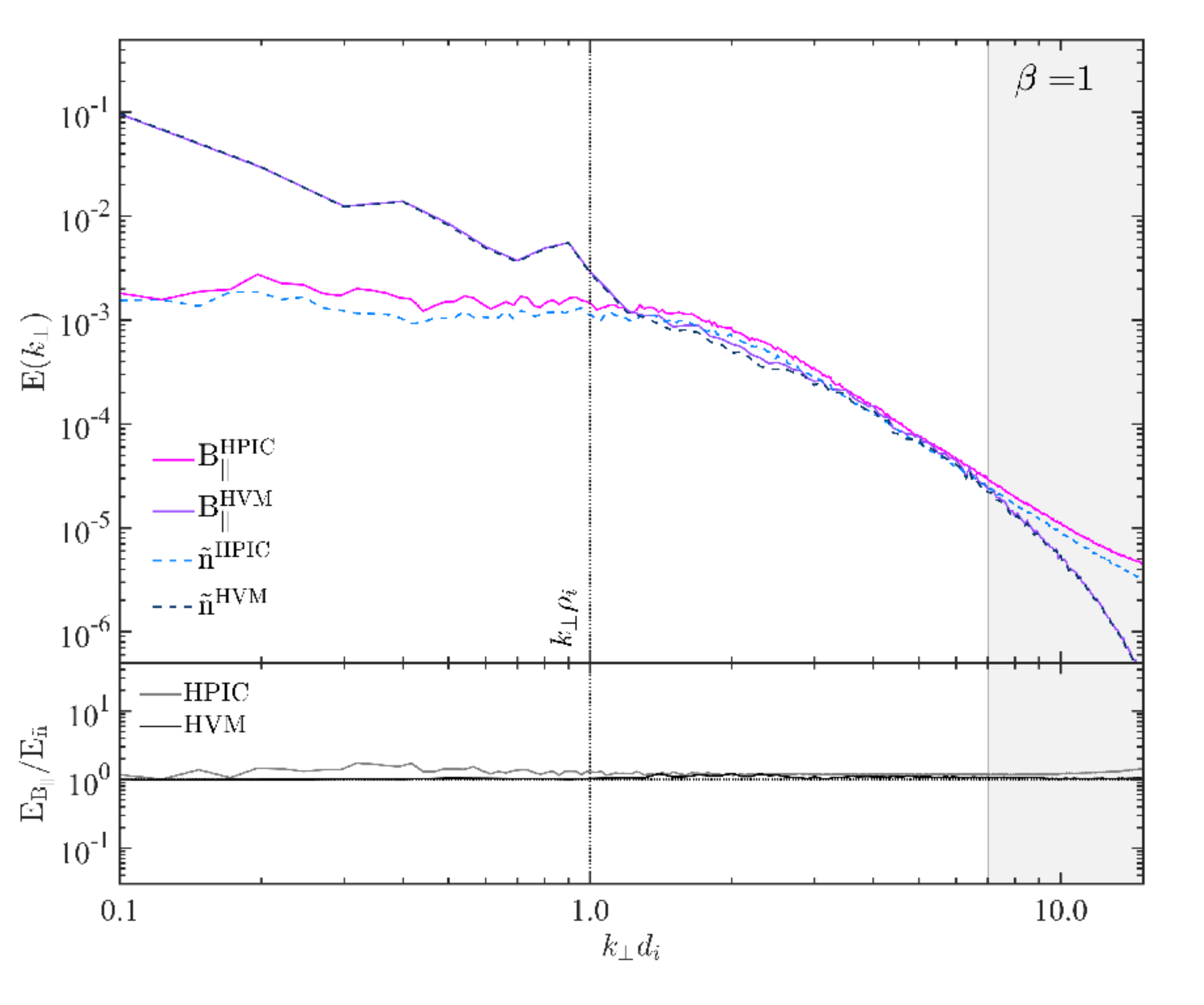}\\
\vspace{-0.35cm}
\includegraphics[width=0.65\textwidth]{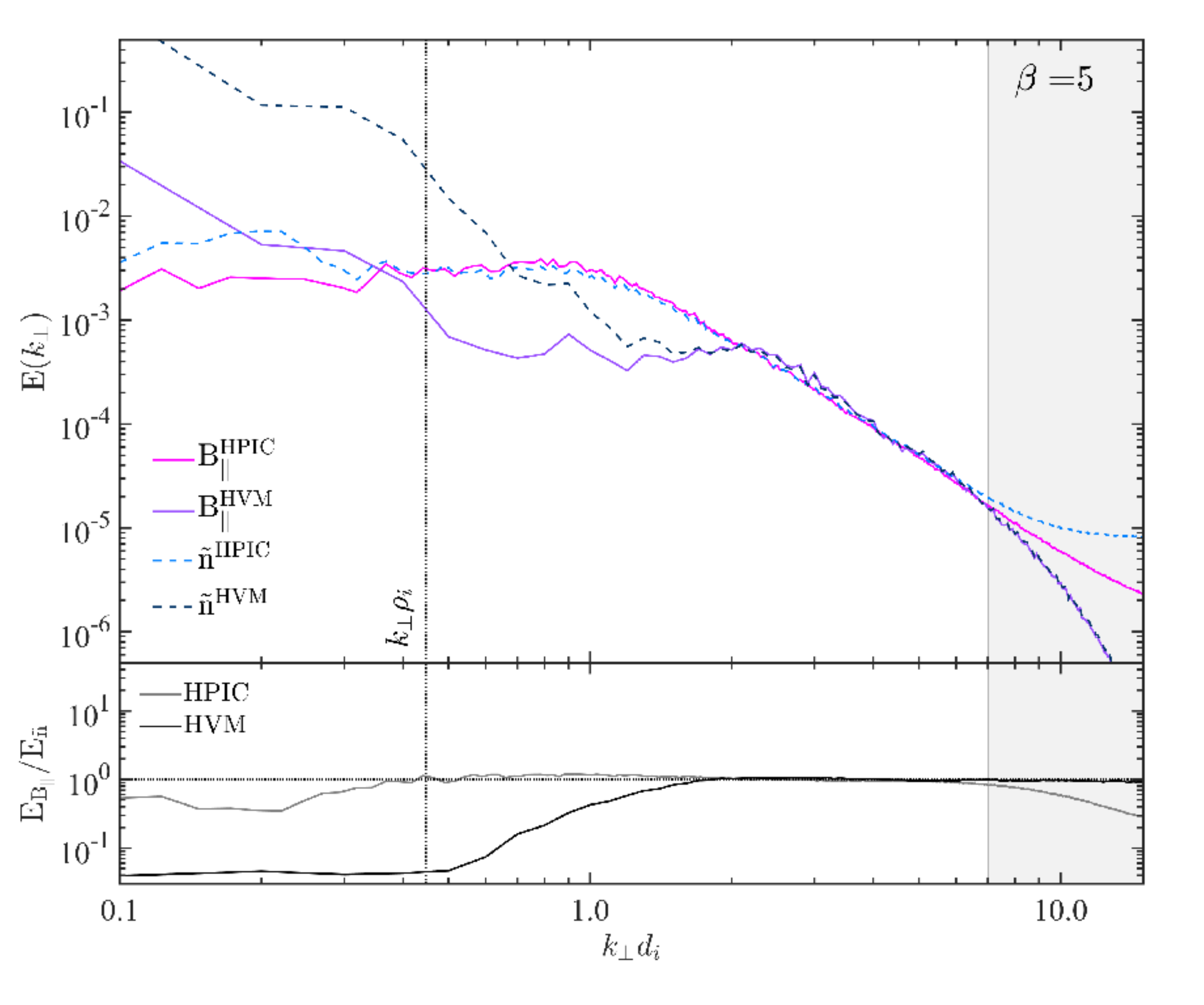}\\
\vspace{-0.35cm}
\caption{Comparison between the HPIC and the HVM spectra of the
  parallel magnetic fluctuations, $E_{B_\|}$ ({\it solid lines}), and of 
  the normalized density fluctuations, $E_{\widetilde{n}}$ ({\it dashed lines}), 
  along with their ratio, $E_{B_\|}/E_{\widetilde{n}}$, in the sub-plots at the bottom,
  for $\beta=0.2$, $1$, and $5$ ({\it top, middle, and bottom panel, respectively}).
  The gray-shaded region marks the range where the spectra are affected by numerical effects.}
\label{fig:spectra_KAW}
\end{figure}
%--------------------------------

In Fig.~\ref{fig:spectra_KAW} we show a comparison between the HPIC and the HVM spectrum of the parallel magnetic fluctuations 
and of the normalized density fluctuations, $E_{B_\|}$ (solid lines) and
$E_{\widetilde{n}}$ (dashed lines), along with their ratio in a sub-plot at the bottom, for the three different plasma beta parameter,
namely $\beta=0.2$, $1$, and $5$ ({\it top, middle, and bottom panel, respectively}). 
Here, the density fluctuations are normalized accordingly with the relation between the density 
and the parallel magnetic fluctuations that is expected for kinetic Alfv\'en waves~\citep{SchekochihinAPJS2009, ChenPRL2013, BoldyrevAPJ2013}, 
%%%%%%%%%%%%%%%%%%%%%%%%%%%%%%%%%%%%%%%%%%%%%%%%%%
\begin{equation}\label{eq:deltan_normaliz}
\delta\widetilde{n}\,=\,\frac{\betai}{2}\,(1\,+\,\tau)\,\delta n\,,
\end{equation}
%%%%%%%%%%%%%%%%%%%%%%%%%%%%%%%%%%%%%%%%%%%%%%%%%%
where $\tau$ is the electron-to-proton temperature ratio 
and it is set to unity for all simulations (the normalization parameter is therefore a function of the plasma beta parameter only).
With such normalization, the prediction for KAW fluctuations would then be $\delta B_\|=\delta\widetilde{n}$, 
so a ratio $E_{B_\|}/E_{\delta\widetilde{n}}$ equal or very close to unity would indicate that 
the main contribution to small-scale turbulence is given by KAW-like fluctuations. 
Such transition to a KAW scenario is supposed to take place at the ion gyro-scale. 
On the contrary, when deviations from unity are significant, it would denote a case in which KAWs fluctuations can only be 
a sub-dominant contribution to small-scale turbulence. 
From Fig.~\ref{fig:spectra_KAW}, we can thus conclude that the $\beta=1$ case ({\it middle panel}) is somehow 
the most Alfv\'enic case, since such ratio is always very close to unity, both at large and at small scales. 
At $\beta=5$, there is instead a clear transition to KAW turbulence as soon as the ion kinetic scales are crossed. 
Note that the scale at which this transition occurs can be affected by the injection, if this takes place at
scales too close to the ion gyroradius, since, as already discussed, 
the system needs to accomplish a self-consistent reprocessing of the large-scale turbulent fluctuations at small scales. 
Nevertheless, even if such scale separation is not completely fulfilled, a transition will anyway take place at a certain scale, smaller than the expected one. 
This is clearly the case of the HVM simulation with $\beta=5$. 
On the contrary, in the HPIC run with the same value of beta, the transition is observed just around $k_\perp\rhoi\sim1$, as expected. 
In this case, the initial injection scale is the same as for the HVM case, i.e., $k_\perp\di\sim 0.3$. 
However, due to the free decay, by the time at which the quasi-steady turbulent state is reached 
that scale has been fully involved in the cascade and thus the effective injection scale has been 
``shifted'' towards larger scales, $k_\perp\di\lesssim 0.1$, i.e., sufficiently far from the ion gyro-scale.
The fact that both the HVM and the HPIC simulations, despite this difference, still reach the same ratio at small scales, 
can be interpreted as a further evidence for the self-consistent response of the plasma system in a defined regime. 
A different result is found instead for $\beta=0.2$, where the ratio $E_{B_\|}/E_{\widetilde{n}}$ 
is significantly different from unity in both cases, 
although the injection is well separated from the ion scales in this $\beta$ regime. 
In particular, a clear and extended plateau at a value $E_{B_\|}/E_{\delta\widetilde{n}}\sim3$ is observed at small scales, starting from $k_\perp\rhoi\sim1$ in the HPIC case and from slightly larger scales in the HVM case.
Such deviation is indeed significant and it thus denotes a sub-dominant contribution of KAWs fluctuations to small-scale turbulence. 
It is worth stressing that, despite the existence of some indications about the presence of magnetosonic/whistler fluctuations in this low-$\beta$ regime, other kinds of fluctuations cannot be excluded a priori. 
Therefore we remind the reader that proving that KAWs are subdominant does not automatically select magnetosonic/whistler fluctuations as the dominant contribution.

\section{Conclusions}\label{sec:Conclusions}

In the present work, we have discussed the main properties of hybrid-kinetic turbulence 
in a collisionless magnetized plasma obtained by means of 2D high-resolution simulations 
ranging from the end of the MHD scales down to scales well below the ion Larmor radius. 
In particular, we have focused our attention on a comparison between two complementary approaches, 
namely (i) hybrid particle-in-cell (HPIC) simulations of freely-decaying Alfv\'enic turbulence and 
(ii) hybrid Vlasov-Maxwell (HVM) simulations of continuously driven partially-compressible turbulence. 
Three values of the plasma beta parameters have been considered, namely $\beta=0.2$, $1$, and $5$, 
corresponding to low-, intermediate- and high-$\beta$ regimes, as observed in the solar wind.
Despite the completely different initialization and injection/drive at the largest scales, 
a very good agreement between the HPIC and HVM simulations has been found at $k_\perp\rhoi\gtrsim1$, 
especially for the turbulent magnetic and density fluctuations. 
In particular, as the ion kinetic scales are approached, the initially different properties 
of the large-scale turbulent fluctuations undergo a self-consistent ``readjustment'' mediated 
by the plasma system, and the same spectral properties are rapidly achieved in the kinetic range. 
A reasonable agreement is also found in the turbulent electric fluctuations, although they are more affected 
by the large-scale injection of momentum fluctuations (through its ideal part, $-\uv_i\times\Bv$) 
in the HVM cases and by the small-scale ppc noise in the HPIC simulations. 
Major differences have been spotted in the velocity fluctuations spectrum, but this feature 
is likely mainly due to the different injection method (free-decay vs. forcing), rather than to 
the different approach (HPIC vs. HVM). In fact, the HVM simulations implement a continuous injection 
of velocity fluctuations that sustains their turbulent cascade, whereas the HPIC let the initial velocity 
fluctuations decay and be dissipated at small scales (as also observed in previous HVM simulations 
of freely-decaying turbulence). The relation between the density and the parallel magnetic fluctuations 
has been also analyzed. A very good agreement between the two approaches has been found. 
In particular, the small-scale turbulent fluctuations have been found to be mainly populated by 
KAW fluctuations for $\beta\geq1$, where a transition is observed around $k_\perp\rhoi\sim1$. 
On the contrary, KAW fluctuations cannot be the main component for $\beta=0.2$, 
where a complete transition never takes place.

Therefore, kinetic-range turbulence in a hybrid-kinetic system seems to be relatively independent 
from the actual injection mechanism or large-scale initial conditions, and from the dissipation 
mechanisms at the smallest scales (numerical damping vs resistivity). 
In fact, whatever large-scale fluctuations one injects, the system will self-consistently ``reprocess'' 
the turbulent fluctuations while they are cascading towards smaller and smaller scales. 
The way in which the system responds to any large-scale injection is indeed mainly dependent 
on the plasma $\beta$ parameter. Despite the limitations due to the reduced model (fluid electrons) 
and to the geometry (2D), the results presented here may have implications for the interpretation of 
SW turbulence data in the context of SW time variability and its different properties at different locations.
In particular, the aspects of kinetic-scale turbulence highlighted in this work 
may prove relevant in the context of the so-called ``Turbulent Dissipation Challenge''~\citep{ParasharJPP2015}, 
such as, for instance, the possible large-scale dependence of the small-scale dissipation and heating in the SW.
\\

The authors acknowledge valuable discussions with A.~Verdini,
L.~Matteini, M.~Velli, W.~Matthaeus, and F.~Pegoraro.  S.S.C. and L.F. are
particularly grateful to the organizers of the {\em``Vlasovia~2016''}
conference, where the ideas leading to this work were born. L.F. is
funded by Fondazione Cassa di Risparmio di Firenze, through the
project ``Giovani Ricercatori Protagonisti''.  L.F. and
S.L. acknowledge the CINECA award under the ISCRA initiative, for the
availability of high performance computing resources and support
(grant HP10BUUOJM). S.S.C. and F.C. acknowledge CINECA (Italy),
for the access to HPC resources where a number of test on the external forcing 
have been performed under ISCRA allocation projects (grants HP10CGW8SW and HP10C04BTP),
and the Max Planck Computing and Data Facility (MPCDF) in Garching (Germany), 
where the HVM simulations were performed.
P.H. acknowledges GACR grant 15-10057S.

%\appendix

%\section{...}\label{app:App1}

% susie put cite commands here, don't bother with citet etc just yet.

\bibliographystyle{jpp}
% Note the spaces between the initials

\end{document}